\documentclass[twocolumn]{aastex}

\usepackage{amsbsy}
\usepackage{amsmath}
\usepackage{textcomp}
\usepackage{amsmath}
\usepackage{color}

\usepackage{graphicx}
\usepackage{epstopdf}
\usepackage{float}

\DeclareMathAlphabet{\mathsfsl}{OT1}{cmss}{m}{sl}

\begin{document}


\title{Magnetic Field Strengths and Grain Alignment Variations in the Local Bubble Wall}

\author{Ilija Medan\altaffilmark{1,2} and B-G Andersson}
\affil{SOFIA Science Center, USRA \\
NASA Ames Research Center, M.S. N232-12 \\
Moffett Field, CA 94035, USA}

\altaffiltext{1}{Physics Department, Santa Clara University, 500 El Camino Real, Santa Clara, CA 95053, USA}
\altaffiltext{2}{Current address: Department of Physics and Astronomy, Georgia State University, Atlanta, GA 30302, USA}

\begin{abstract}

Optical and infrared continuum polarization from the interstellar medium is known to generally be due to irregular dust grains aligned with the magnetic field. This provides an important tool to probe the geometry and strength of those fields, particularly if the variations in the grain alignment efficiencies can be understood. Here, we examine polarization variations observed throughout the wall of the Local Bubble, using a large polarization survey of the North Galactic cap (\textit{b}$>30^\circ$) from \citet{berdyugin2014}. These data are analyzed together with archival photometric and spectroscopic data along with the mapping of the Local Bubble by \citet{lallement2003}. We can model the observational data by assuming that the alignment driving mechanism is due to the radiation from the surrounding star field.  In particular we find that the fractional polarization is dominated by the light from the OB associations within 150 pc of the sun, but is largely insensitive to the radiation field from red field stars.  This behavior is consistent with the expected wavelength dependence of radiative grain alignment theory.  We also probe the relative strength of the magnetic field in the wall of the Local Bubble using the Davis-Chandrasekhar-Fermi method. We find evidence for a systematically varying field strength distribution, where the variations in the field are correlated with the variations in grain alignment efficiency, indicating that the relatively higher field strength regions might represent a compression of the wall by the interaction of the outflow in the Local Bubble and the opposing flows by the surrounding OB associations.

\end{abstract}


\section{Introduction} \label{sec:intro}
Interstellar polarization has been known, almost since its discovery, to be due to asymmetric dust grains aligned with the magnetic field \citep{hiltner1949a,hiltner1949b}.  This means that the polarization traces the large scale structure of the plane-of-the-sky magnetic field, both locally \citep[e.g.,][]{berdyugin2014}, and on galactic scales \citep[e.g.,][]{scarrott1996}.  In addition, because the magnetic field-lines in a plasma, \textit{de facto}, behave as tensed strings \citep[e.g.][]{choudhuri1998}, the local variations in the polarization direction can be used as a probe of the magnetic field strength, if the gas turbulence and density can be estimated \citep{davis1951b, chandrasekhar1953}.  Dust-induced polarization can be observed with relative ease in material of almost any opacity, either due to dichroic extinction polarization in the ultraviolet through near infrared wavelength range, or in the far infrared through mm-wave range due to dichroic emission. However, a fully quantitative understanding of the relationship between observed polarization and magnetic field characteristics is still lacking.  

Over the last few decades a quantitative theory of grain alignment has been developed \citep[e.g.][]{dolginov1976,draine1996,lazarian2007} which provides a number of specific, testable, predictions.  In this ``Radiative Alignment Torque" (RAT) theory, anisotropic radiation spins the grains up and, if their bulk is paramagnetic, aligns them with the magnetic field \citep{lazarian2007}.  While a significant number of theoretical predictions from RAT alignment have now been qualitatively confirmed \citep{bga2015b}, the details, including specific coupling parameters (e.g. the alignment time for a specific grain exposed to a given radiation field), still need to be better constrained for the theory to be fully predictive in practice and to allow detailed forward modeling. 

Because the local distributions of both the interstellar medium (ISM) and bright stars are well studied \citep{lallement2003,dezeeuw1999}, the local interstellar medium provides an important laboratory for studying grain alignment and dust-induced polarization.

The immediate Galactic vicinity of the Sun is dominated by a low density, ionized structure, commonly referred to as the ``Local Bubble''.  It is bounded by relatively higher density material as traced by Sodium and Calcium absorption line measurements, as well as extinction data \citep{lallement2003,welsh2010,lallement2014}.  Such measurements show a roughly cylindrical structure with a typical radius of about 100-175 pc, with missing ends towards the north and south Galactic poles.   This structure is generally interpreted as being due to strong stellar winds and supernovae evacuating the space, with ``blow-outs" in the directions out of the Galactic plane \citep{lallement2003}.

An on-going debate with respect to the Local Bubble is its thermodynamic state. Some extreme ultraviolet (EUV) and soft X-rays observations, as well as those of highly excited, and/or highly ionized states of several elements, suggest a hot state with high thermal pressure of $\sim15,000-20,000$ K cm$^{-3}$ \citep{bowyer1995,snowden1998,welsh2009}. In contrast, measurements using the fine-structure transitions of \ion{C}{1} \citep{jenkins2002} find a substantially lower thermal pressure inside the bubble. A complementary probe of the state of the Local Bubble is the pressure in the surrounding wall, which can be inferred from the magnetic field strength. To accomplish this, \citet{bga2006} used optical R-band polarization in the direction of the Southern Coalsack at \textit{l,b}$\approx$300,0, towards stars with known distances, to carry out a Davis-Chandrasekhar-Fermi analysis.   Complementing the polarization with high resolution observations of the ultraviolet fine-structure lines of C I  towards $\mu^2$ Cru, and $H_2$ analysis from \citet{lehner2003} to estimate the gas pressure, density and turbulence, they derived a magnetic field strength of B$_\perp = 8^{+5}_{-3}  \mu$G, equivalent to a magnetic pressure of P$_B/k\approx18,000$ K cm$^{-3}$, consistent with the results from the X-ray and the EUV observations.

The present study is, in part, aimed at extending this earlier result to other areas of the Local Bubble wall. We also intend to use the polarization measurements to further test our understanding of grain alignment physics. We take advantage of the detailed mapping of the Local Bubble presented by \citet{lallement2003} and the extensive survey of polarimetry towards near-by stars at high Galactic latitudes presented by \citet{berdyugin2014}.  Combining these with stellar classifications from \citet{wright2003}, and B$_T$,V$_T$ photometric data from \citet{tycho2000} and J,H,K$_S$ data from \citet{cutri2003} allows a  study of both the magnetic field and the grain alignment properties over a large part of the Local Bubble wall.

The quantitative relationship between an observed level of polarization and the strength of the magnetic field depends on a number of grain characteristics, including shape, mineralogy and internal temperature, as well as grain alignment efficiencies, both driving and damping the alignment and the internal ISM structure \citep{lazarian2015}, including turbulence \citep{jones1992}.  Grains in most grain alignment theories rotate around the magnetic field lines, implying that dust-induced polarization can only probe the magnetic field components perpendicular to the line of sight.  However, by studying a coherent, yet varied, system at moderate opacities, such as the Local Bubble wall, we can address many aspects of grain alignment theory and the interrelation between polarization and magnetic fields.

The rest of the paper is organized as follows:   The polarization survey used for the analysis is described in Section \ref{sec:polar}, the archival photometric and spectroscopic data are described in Section \ref{sec:extinct}, and basis for the Local Bubble wall geometry employed is described in Section \ref{walldist}. Section \ref{subsec:distances} argues  that the stellar sources analyzed only probe the Local Bubble wall and not also more distant extinction screens.

Section \ref{fracpol} discusses the analysis in terms of the observed fractional polarization.  For this part, Section \ref{the grains} describes our assumptions concerning the gas and grain characteristics within the wall of the bubble and Section \ref{turb} discusses how the turbulence throughout the probed part of the Local Bubble wall is found to be roughly constant.  Section \ref{disalign} describes the modeling of grain alignment from the observed fractional polarization.  Section \ref{spatial_variations} analyzes the spatial variations found in the fractional polarization in terms of the projection of a large-scale local field (Section \ref{projection_sec}) and as due to radiative alignment (Section \ref{radiation_sec}).

Section \ref{sec:pol_ang_disp} describes the analysis of the position angle dispersions, used as the basis for a Davis-Chandrasekhar-Fermi analysis.

Section \ref{sec:results} provides the results of the analysis of the grain alignment variations (Section \ref{subsection:results,grain_alignment}), and the positional angle dispersion (Section \ref{subsection:results,dispersion}). Section \ref{sec:discussion} provides a discussion of the implications of our results. Finally a brief summary and our conclusions are given in Section \ref{sec:conclusions}.
 
\section{Data Analysis} \label{sec:analysis}

\subsection{Polarimetry}\label{sec:polar}

The polarization survey by \citet{berdyugin2014} primarily covers the sky at Galactic latitudes \textit{b}$>30^\circ$, which therefore is the focus of our analysis. Because of the asymmetric error characteristics of polarimetry data \citep{vaillancourt2006}, only stars with $p\geq2.5\sigma_p$ were used in our analysis.    To complement the polarization data, we cross-referenced the target stars in \citet{berdyugin2014} with the catalog of spectrally classified stars in the Tycho database by \citet{wright2003}, and extracted  photometry from the Tycho \citep{tycho2000} and 2MASS \citep{skrutskie2006} surveys.  Trigonometric distances were extracted from, in order of priority, the second data release from the Gaia mission \citep{gaiaDR2}, the first data release from the Gaia mission \citep{gaiaDR1} and the Hipparcos catalog \citep{hip1997}.  To overcome many of the unknowns in the polarization on individual lines of sight, our analysis is then performed on the average behavior over sub-regions of the Local Bubble wall.

We divided the sky covered by the \citet{berdyugin2014} sample into 15$^\circ$ bands in Galactic longitude.  To form regions of similar solid angle, we defined three Galactic latitude bands of $30-41.8^{\circ}$, $41.8-56.4^{\circ}$ and $56.4-90^{\circ}$.  This binning provides an adequate number of samples per region, while allowing for some angular resolution.

Because most of the observations in the \citet{berdyugin2014} sample were acquired in ``white light", a slightly increased uncertainty exists in the variations of the polarization, compared to measurements in a well defined spectral band.  This uncertainty is due to the varying weighting of the wavelength-dependent polarization curve for different spectral classes (over the response function of the CCD) and the known variations of the polarization curve with varying extinction \citep{bga2007}.  Since the spectral classes of the stars show no correlation with distance or extinction, the former should, at worst, introduce additional random error in our analysis.  The latter should also not cause any systematic errors given the small range of visual extinctions probed (see Table \ref{summarytab}).  Becasue the position angles of interstellar polarization generally do not vary significantly with wavelength \citep{whittet1992}, we do not expect any additional uncertainty to be introduced in the polarization angle analysis.
  
\subsection{Visual Extinctions}\label{sec:extinct}

To calculate the visual extinctions of the targets, B$_T$,V$_T$ and J,H,K$_S$ photometry \citep{tycho2000,skrutskie2006} were transformed to the standard extended Johnson system \citep{mamajek2002,carpenter2001}. Intrinsic colors were estimated based on the spectral classifications given in \citet{wright2003} and tables of intrinsic colors in \citet{cox2000}.  

For stars without luminosity classification, we assumed luminosity class V for spectral classes earlier than G5, and III for later-type stars. This assumed classification is based on two lines of reasoning.   
First, \textbf{the} Henry Draper (HD) catalog constitutes an all-sky magnitude limited sample (m$_{ph}<8$ with many regions and the extension catalog going at least a magnitude deeper \citep{shapley1925}), with an apparent magnitude distribution similar to the \citet{berdyugin2014} sample.  The Michigan spectral classification survey of the HD catalog  \citep{michiganV1,michiganV2,michiganV3,michiganV4,michiganV5} contains $>$180,000 stars.  Of the stars classified as Main Sequence, 87\% are earlier than G5.  Correspondingly, 86\% of the giants are classified as later than G4. 
Second, main sequence stars later than G5 beyond the distance of the Local Bubble wall would generally be too faint to be detected in the present sample.

Having established the optical and near infrared color excesses, we used the relationship from \citet{whittet1978} to calculate the total-to-selective extinction R$_V$:
\begin{equation}
R_V=1.1\cdot\frac{E_{V-K}}{E_{B-V}}
\end{equation}
which then yields the visual extinction through: $A_V=R_V\cdot E_{B-V}$.  Of the 2,400 stars in the original sample, the combined data sets allowed reliable ($>2\sigma$) visual extinctions to be estimated for 1,066 stars.

\subsection{Wall Distance and Orientation} \label{walldist}

\begin{figure*}
    \epsscale{1.1}
    \plotone{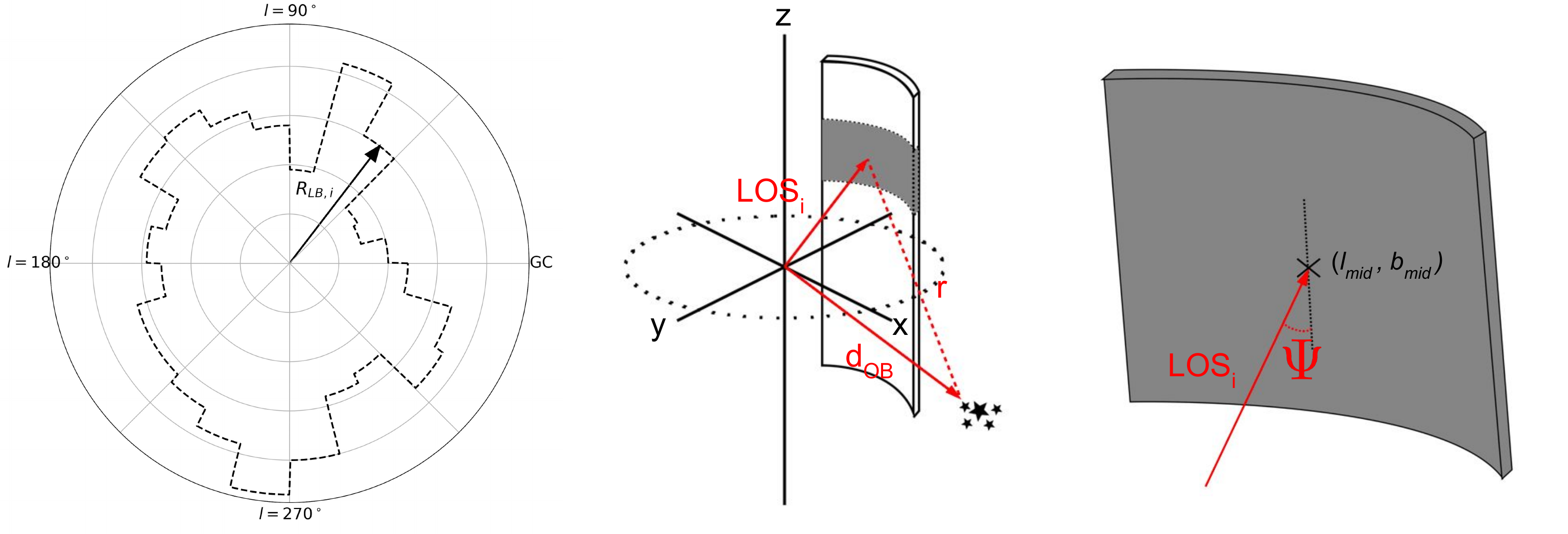}
    \caption{(Left) The Local Bubble as described by our model, where the distance to the wall is assumed to be constant for each of the regions in the study. The the Earth is located at the center.  The radius for one of the regions is indicated by the arrow. (Center) A diagram of one of the wall segments.  The grey section in the segment is one of the regions of our study. The arrow labeled ``LOS" represents the line of sight to the center of the region with length  R$_{LB,i}/cos(\langle b_i\rangle)$.  The arrow labeled ``$d_{OB}$" represents the vector to a star or cluster of stars beyond the wall, and ``r" is the distance between the center of the region and a star or cluster of stars. (Right) A diagram of one region in our study. The line of sight vector intersects the center of the region. The angle between the line of sight vector and the wall in this region is the wall inclination for the region and is labeled as ``$\Psi$".}
   \label{lb_diagram}
\end{figure*}

Based on the general shape of the Local Bubble, we approximate it as a set of cylindrical slices with a fixed radius for each 15$^\circ$ longitude segment, as shown in the left panel of Figure \ref{lb_diagram}.  For each longitude segment, we estimated the Local Bubble radius as the average in the Galactic Plane using Figure 4 in \citet{lallement2003}.  The global average radius from these estimates is about 150 pc.

The Local Bubble wall distance for each (\textit{l,b}) region was then estimated as d=R$_{LB}/cos(\langle b\rangle)$, where R$_{LB}$ is the estimated bubble radius for the longitude segment and $\langle b\rangle$ is the Galactic latitude of the region, as shown in the center panel of Figure \ref{lb_diagram}.  At each height, the region making up the wall is also assigned a slope, which is \textbf{equivalent to} the angle the wall makes with the line of sight, as shown in the right panel of Figure \ref{lb_diagram}.

In order to estimate the average wall orientations ($\Psi$; Table \ref{summarytab}) for each of the regions in our study, we used the vertical cross sections plotted in Figures 7 and 8 of \citet{lallement2003}), interpolating, as required, to estimate the wall angle at the midpoint of our regions. 

\subsection{Extinction Distances} \label{subsec:distances}

Because of the limited opacity in the local ISM and the vector nature of polarization, it is important to evaluate the existence of multiple extinction layers, probed by the stars in our sample. This is because non-aligned magnetic fields in such separate layers could introduce systematic errors in the analysis. 

\begin{figure}[t]
	\epsscale{1.2}
	\plotone{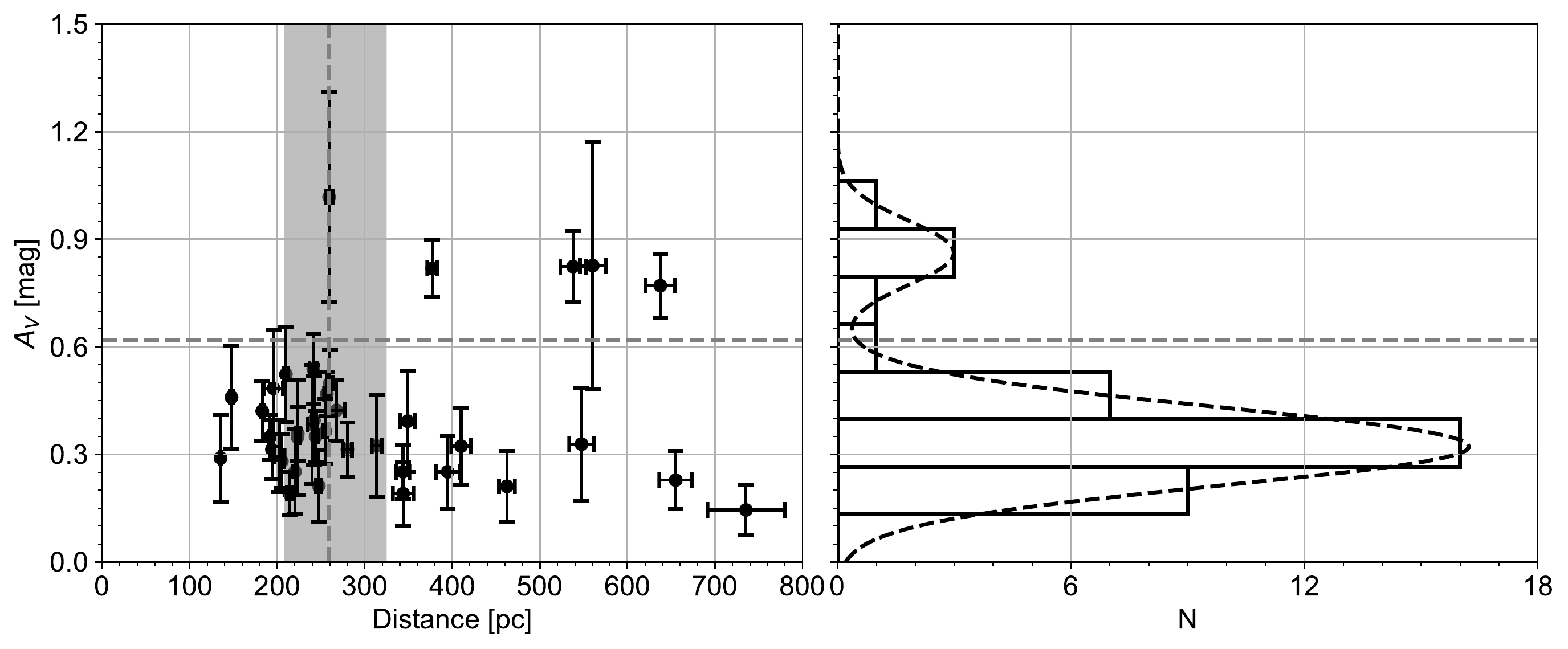}
	\caption{Example of the distribution in extinction (right panel, solid black line), corresponding two component fit (right panel, dashed black line), division determination (both panels, horizontal dashed gray line) and determined extinction screen distance (left panel, vertical dashed gray line) for \textit{l}$=315-330^{\circ}$. The solid gray box in the left panel represent the spread in Local Bubble wall distance for this region from \citet{lallement2003}, where the spread shows the change in wall distance along the line of sight between \textit{b}=$30-56.4^{\circ}$}
	\label{stepexample}
\end{figure}

Because of the limited number of stars in each region, it is often not possible to conclusively disprove the existence of additional extinction layers, as the sample may not uniformly probe the distance space.  Often an inhomogeneous screen (i.e. a clumpy medium) can have properties similar to multiple extinction screens, as shown in Figure \ref{stepexample}.  However, the new, improved Gaia distances allow better estimates than previously possible. 

To probe for multiple extinction layers, we examined the distributions of extinction and polarization measurements for each of the 15$^\circ$ longitude cuts. For this analysis, and to ensure numerical stability, we included only sources with distance measurements with S/N$>2$ and we excluded a small number of outlier sources ($<2\%$) with extinction or polarization measurements more than three standard deviations from the average for the region.  

We fit the A$_{V}$ and p distributions for each region with one- and two- component Gaussian functions. For fits favoring two components, with means separated by more than 3$\sigma$ (mutual uncertainty), a dividing line between the two populations was defined as the midpoint between the means (Figure \ref{stepexample}, horizontal dashed gray lines). The distance to the second screen, was then taken to be the distance to be the nearest star in that ``upper" sub-sample (Figure \ref{stepexample}, vertical dashed gray line).  The results are shown in Table \ref{extdistsum}. If no increase is listed in Table \ref{extdistsum}, then a one-component fit was favored. As the net polarization and extinction before the wall distance is very small \citep{lallement2003}, the extinction and polarization seen in our data is assumed to be due to the Local Bubble wall (or material behind it).  The distance to, and extent of, the Local Bubble wall on small scales is uncertain enough that some stars could appear on the near side of the estimated bubble wall distance (e.g. Figure \ref{stepexample}) even when behind or embedded in the wall. In all regions, we find that the second sub-sample has a distance equal to, within the uncertainties, the distance to the Local Bubble wall as derived from the \citet{lallement2003} data.  Therefore, the sub-samples of higher average extinction and/or polarization are most likely indicative of a clumpy density structure in the Local Bubble wall, rather than additional, physically distinct, extinction layers along the line of sight.  In the following analysis, we will therefore assume a single extinction and polarization screen for all map bins.

\begin{figure}[t]
\begin{deluxetable}{lccc}
	\tablecaption{Summary of estimated polarization and extinction increases and assumed Local Bubble wall distances. \label{extdistsum}}
	\tablehead{
		\colhead{\textit{l}} & \colhead{Wall Distance\tablenotemark{$\dag$}} & \colhead{$p$ Step}  & \colhead{$A_V$ Step}\\
		\colhead{[$^{\circ}$]} & \colhead{[pc]} & \colhead{[pc]} & \colhead{[pc]}}
	\startdata
	0-15    & $134^{+47}_{-19}$  & \nodata  & \nodata  \\
	15-30   & $101^{+35}_{-14}$  & \nodata  & \nodata \\
	30-45   & $107^{+38}_{-15}$  & \nodata  & \nodata \\
	45-60   & $201^{+70}_{-28}$  & \nodata  & \nodata \\
	60-75   & $282^{+97}_{-40}$  & $229$  & \nodata \\
	75-90   & $127^{+45}_{-17}$  & \nodata  & \nodata \\
	90-105  & $188^{+65}_{-26}$  & \nodata  & \nodata \\
	105-120 & $215^{+74}_{-30}$  & \nodata & $278$ \\
	120-135 & $241^{+84}_{-33}$  & $261$ & \nodata \\
	135-150 & $235^{+81}_{-33}$  & $283$ & \nodata \\
	150-165 & $174^{+61}_{-24}$  & \nodata  & \nodata \\
	165-180 & $195^{+67}_{-28}$  & \nodata & \nodata \\
	180-195 & $174^{+61}_{-24}$  & \nodata & \nodata \\
	195-210 & $215^{+74}_{-30}$  & \nodata  & \nodata \\
	210-225 & $215^{+74}_{-30}$  & $169$ & \nodata \\
	225-240 & $228^{+79}_{-32}$  & \nodata & \nodata \\
	240-255 & $255^{+88}_{-36}$  & \nodata & \nodata \\
	255-270 & $315^{+110}_{-44}$ & \nodata & \nodata \\
	270-285 & $268^{+93}_{-37}$  & \nodata & \nodata \\
	285-300 & $188^{+65}_{-26}$  & \nodata & \nodata \\
	300-315 & $174^{+61}_{-24}$  & \nodata & \nodata  \\
	315-330 & $241^{+84}_{-33}$  & \nodata & $259$ \\
	330-345 & $228^{+79}_{-32}$  & \nodata & \nodata \\
	345-360 & $161^{+56}_{-22}$  & \nodata & \nodata
	\enddata
	\vspace{3mm}
	\tablenotetext{$\dag$}{Because most regions in the upper most \textit{b} band ($56.4-90^{\circ}$) consist of blow-out regions, we consider the spread of wall distances between the lower two bands. The wall distance listed then is the estimated wall distance at \textit{b}=$41.8^{\circ}$ and the spread listed is the change in wall distance along the line of sight between \textit{b}=$30-56.4^{\circ}$}
	\tablecomments{If an increase distance is not listed, this means, according to our criteria, no secondary increase was observed}
\end{deluxetable}
\end{figure}

\subsection{Fractional Polarization} \label{fracpol}

Not all dust grains contribute to observed polarization.  As noted in the introduction, this can be due to alignment efficiency variations (both driving and damping mechanisms), as well as grain shape (spherical grains can't induce polarization) and composition (carbonaceous grains are not susceptible to most alignment mechanisms \citep{bga2015b}).  

Also, because the aligned grains are spinning around the magnetic field lines, dust-induced polarization can only probe magnetic field components perpendicular to the line of sight. Because polarization is a vector entity, while areas of parallel magnetic fields along the line of sight will cause increased observed polarization, crossed magnetic field directions along the line of sight will cause a decreased polarization. 

Hence, intrinsic physics (wavelength dependent alignment efficiency, grain size and shape variations, etc.) line of sight effects (line of sight projection of the local magnetic field as well as its variations along the line of sight) and observational constraints (target selection and sample, distances, etc.) affect the data and their interpretation. The following subsections quantify some of these considerations\textbf{.} 

\subsubsection{Grain and Gas Characteristics} \label{the grains}

For this study, we will assume that the grain size, shape and mineralogy distributions are fixed.  Determining the space density and temperature of the gas requires e.g. H$_2$ excitation data \citep{lehner2003} or similar observations, which we do not have for the current lines of sight.  We will therefore initially assume that also the gas density and temperature are constant throughout the Local Bubble wall, but will return to this assumption where warranted (e.g. Section \ref{alpha}).  The indication of clumpyness in the wall, discussed above (Section \ref{subsec:distances}), is limited enough that it should not significantly affect the analysis.

\subsubsection{Influence of Geometry and Turbulence} \label{turb}

Theoretical models of super-bubble structure \citep{stil2009} indicate that the magnetic field tends to follow the direction of the wall.  We can therefore account for large-scale projection effects in the Local Bubble wall by estimating the average direction of the magnetic field, using the wall inclination relative to the line of sight, as described above (Section \ref{walldist}). 

\begin{figure}[t]
	\epsscale{1.2}
	\plotone{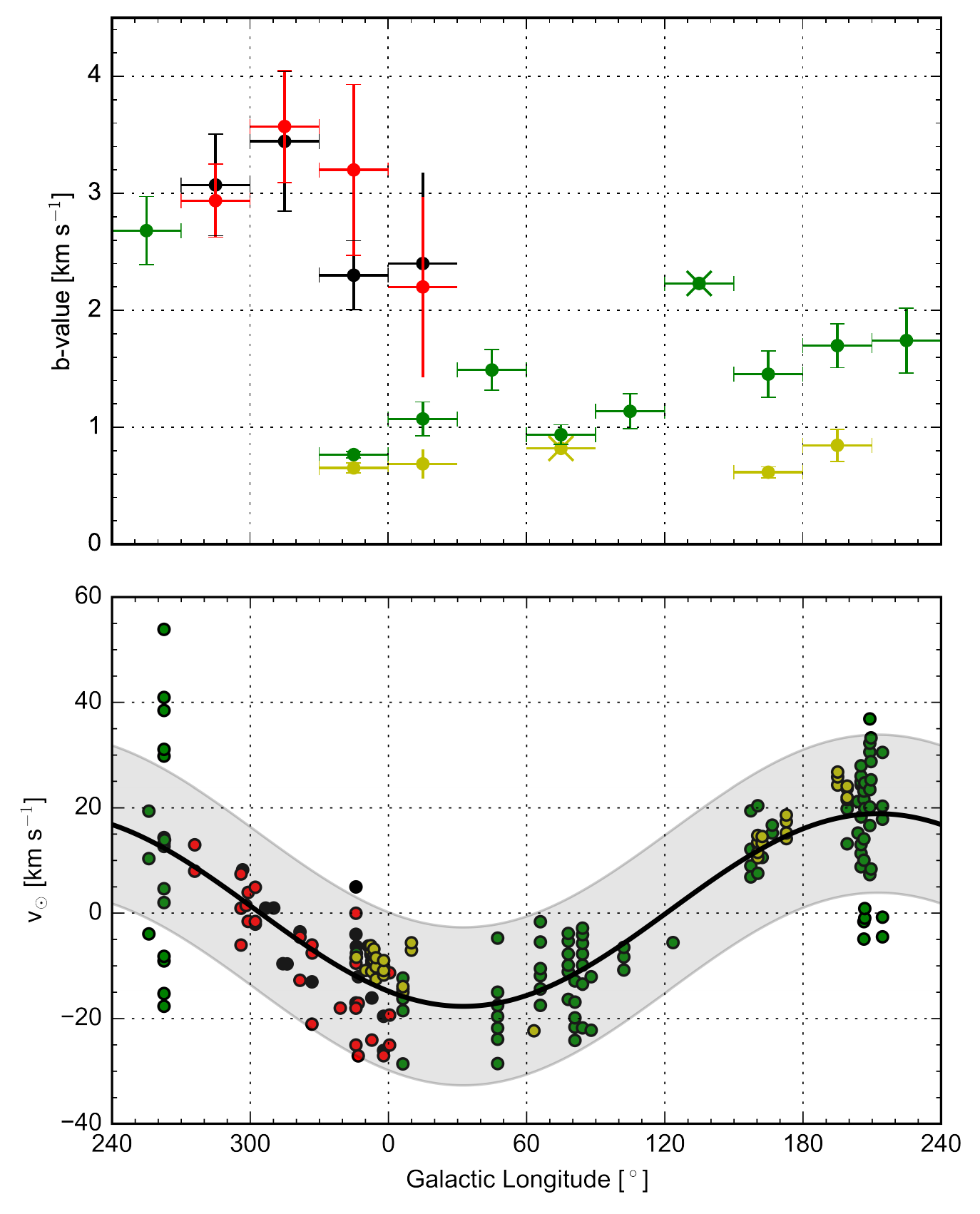}
	\caption{Upper Panel: Averaged b-values as a function of Galactic Longitude from \citet{welty2001} (yellow), \citet{welty1996} (green) and \citet{crawford1991} (black, \ion{Na}{1}; red, \ion{Ca}{2}). All longitude bin widths for the averages are $30^\circ$. Bins with an ``X" through them indicate a bin with only one b-value that met our criteria. \\ Lower Panel: Heliocentric velocity of the absorption line components as a function of Galactic Longitude where the colors correspond to those of the upper panel.}
	\label{turb_plot}
\end{figure}

To constrain the depolarization effect by changes in the direction of the magnetic field, due to turbulence along the line of sight within the Local Bubble wall, we used high-resolution optical absorption line surveys of alkali elements.  We extracted b-values for \ion{K}{1} absorption line data from \citet[spectral resolution of $\delta v\sim0.4-1.8\ km\ s^{-1}$]{welty2001}, \ion{Ca}{2} data from \citet[$\delta v\sim0.3-1.2\ km\ s^{-1}$]{welty1996}, and for \ion{Na}{1} and \ion{Ca}{2} data from \citet[$\delta v\sim3.6\ km\ s^{-1}$]{crawford1991}, which together cover most Galactic longitudes (Figure \ref{turb_plot}).  To match the absorption line data as closely as possible to the Local Bubble wall gas, we selected those stars located at distances between 100 and 500 pc. Additionally, we only included absorption lines with $N\geq 5\times 10^{10}cm^{-2}$ for the respective spectral lines.  \citet{jenkins2011} used ultraviolet observations of the fine structure lines of C I to show that the Mach number for diffuse cold neutral gas is typically in the range 1-4.  The ionization potentials of \ion{K}{1} and \ion{Na}{1} are lower than that for \ion{C}{1} while those of \ion{Ca}{2} and \ion{C}{1} are similar.  Because the sound speed (for an ideal gas) rises with temperature, we will therefore here assume that the observed line widths are dominated by the turbulence and that the thermal component can be ignored.

As shown by e.g. \citet{welty2001} or \citet{bga2002a}, neutral gas in the ISM can have line widths well below 1$\ km\ s^{-1}$ with similar component separation.  Therefore, even at the resolution of the \citet{crawford1991} data, the instrumental resolution imposes severe limitations on determining the intrinsic line widths of individual components. The b-values from such lower resolution data should therefore be treated as upper limits.

To facilitate a comparison of the spatial variations of turbulence using the polarization data, we averaged the b-values in longitude bins of 30$^\circ$.  The averaged b-values are shown in Figure \ref{turb_plot}. The measured line widths are roughly constant with longitude, with a possible enhancement in the fourth quadrant (where the measurements \textbf{are} predominantly from lower resolution data). This result indicates that the depolarizing effects of non-parallel magnetic fields within the Local Bubble wall is not highly variable by location, and therefore that the variation  in p/A$_V$ is dominated by grain alignment rather than gas turbulence \citep[cf.][]{jones1992}.

\subsubsection{Fractional Polarization Model} \label{disalign}

Several methods have been employed to evaluate the grain alignment efficiency, primarily the use of the fractional polarization, p/A$_V$\footnote{We will distinguish between the measured value of the relative amount of polarization and visual extinction, p/A$_V$, which we will refer to as the ``fractional polarization", and the intrinsic fraction of polarization to column density (``polarization efficiency"), which is a theory/modeling concept.} \citep[eg][]{jones1989,whittet2008,cashman2014}, and the location of the peak of the polarization curve \citep[eg][]{whittet2001,bga2007}. Whereas the former, as a probe of alignment efficiency and especially for individual targets, is susceptible to modification by line of sight turbulence \citep{jones1992} and the orientation of the field, it has the advantage of only requiring single-band polarimetry.  

In the presence of turbulence, the fractional polarization as a function of visual extinction is expected \citep{jones1992} to follow a power law relation of the form:

\begin{equation}
\frac{p}{A_V}=\beta\, A_V^{-\alpha}
\label{powerlaw}
\end{equation}

In such a relationship $\alpha$ depends on the turbulence of the material (and may, in addition, depend on grain alignment variations along the line of sight \citep{alves2014,jones2015}). Without variations in the alignment efficiency, $\alpha=-0.5$ is characteristic of a fully turbulent medium \citep{jones1992}.  The parameter $\beta$ -- equivalent to p/A$_V$(A$_V$=1) -- is sensitive to the fraction of alignable grains, the grain alignment efficiency, and the orientation of the field relative to the line of sight.  

\begin{figure}[t]
	\epsscale{1.2}
	\plotone{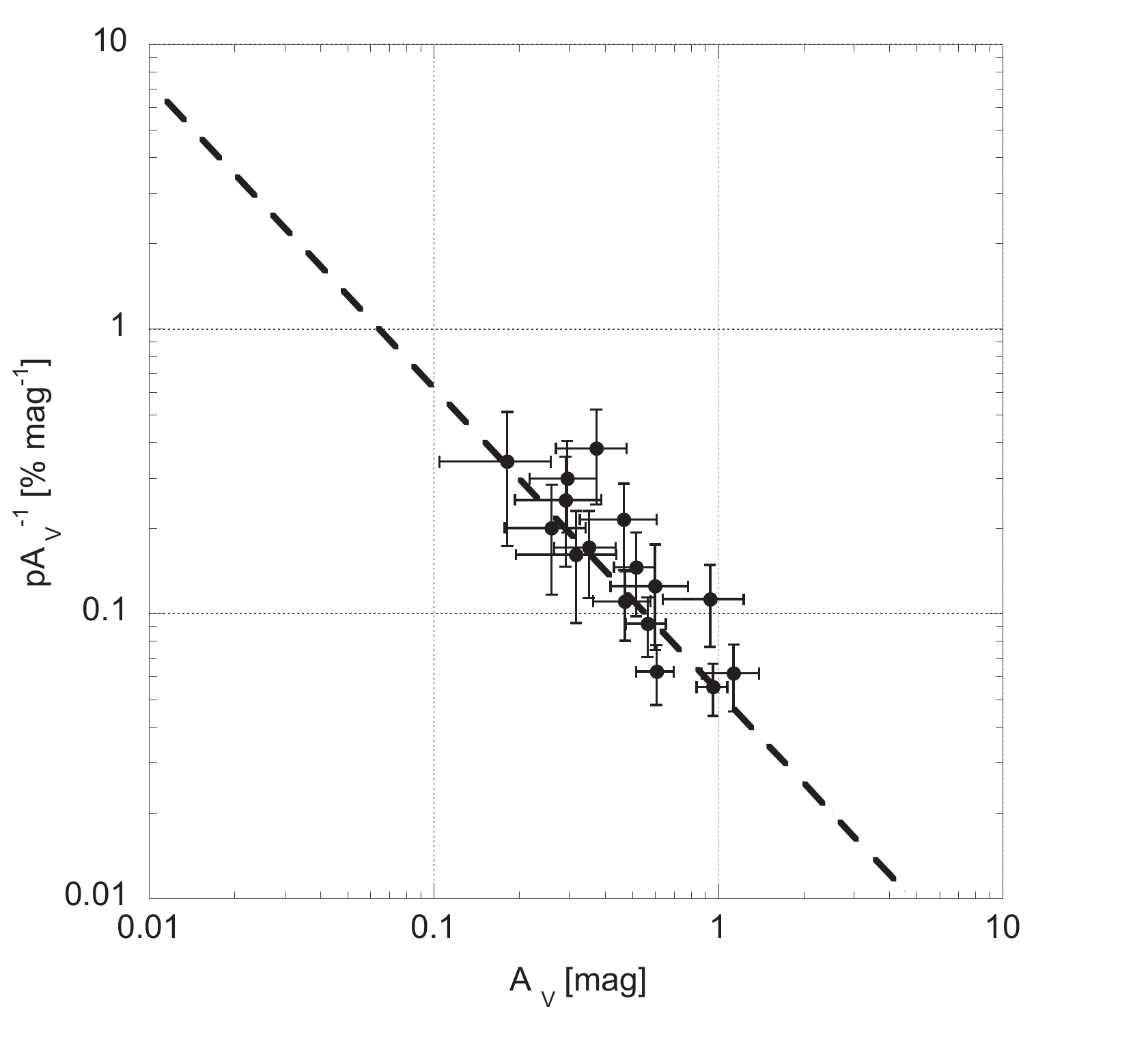}
	\caption{The fractional polarization and visual extinction for stars within \textit{l, b}$=210-225^\circ$,  $30-41.8^\circ$ plotted on a logarithmic scale. The least-squares power-law fit of $\alpha=-1.07\pm0.18$  and $\beta=0.0533\pm0.0180$ is plotted as the dashed line.}
	\label{powerindexplot}
\end{figure}

Figure \ref{powerindexplot} shows the resulting fit of equ. \ref{powerlaw} to the data for one of our regions.  The best-fit parameters for each area are listed in Table \ref{summarytab}.  For those regions where insufficient data exist to provide meaningful fits, no parameters are listed.

To account for any dependence of the measured fractional polarization on the derived wall inclination ($\Psi$; right panel of Figure \ref{lb_diagram}), and thus projection of the average magnetic field direction, we use $\beta (sin\Psi)^{-1}$ as our modified probe of the grain alignment efficiency, to be compared to model predictions (c.f. \citet{axon1977}).

An expected exception to the validity of this projection argument is for those regions where the mean direction of the local magnetic field is close to the line of sight.  This is primarily a concern in the ``blow-out regions" of the Local Bubble, where the bubble wall has been mostly removed by mechanical motion.  In those regions, the average field direction should be close to the line of sight, and the fractional polarization will not probe the average field orientation.  We do not, however, expect to observe zero polarization even in such regions, as the turbulence will cause a component of the field to be perpendicular to the line of sight, probing the magnetic field dispersion.
	
To estimate the minimal $\Psi$ where the projected field direction can be reasonably estimated, we use the polarization dispersion found by \citet{bga2006} for the bubble wall towards \textit{l,b}$\sim$300,0 of $\Delta\theta\approx26^\circ$.  Based on this result, we would expect that the turbulence dominates the effects of projection of the average field direction for $\Psi\sim13^\circ$.  We therefore apply the $sin(\Psi)^{-1}$ modification to $\beta$ only for regions with $\Psi>13^\circ$.  This choice is supported by the results for regions with low $\Psi$ where we find uniformly low $\beta$ values  ($\overline{\beta}=0.15\pm0.04$ for $\Psi\leq13^{\circ}$). Using a Student t-test for $\beta$ values with $\Psi\leq13^{\circ}$ and $\Psi>13^{\circ}$, we find that $\overline{\beta}$ is significantly lower for $\Psi\leq13^{\circ}$ compared to $\Psi>13^{\circ}$ (t=-2.67 and p=0.0138).  We discuss the impact of this modification on our modeling in section \ref{radtoOB}.

\subsection{Spatial Variations in the Fractional Polarization} \label{spatial_variations}

The $\alpha$ parameter (equ. \ref{powerlaw}) yields an average value of $\langle\alpha\rangle=-0.82\pm0.06$ (error on mean), with no clear, systematic, spatial variations.

\begin{figure*}[t]
	\centering
	\epsscale{1.1}
	\plottwo{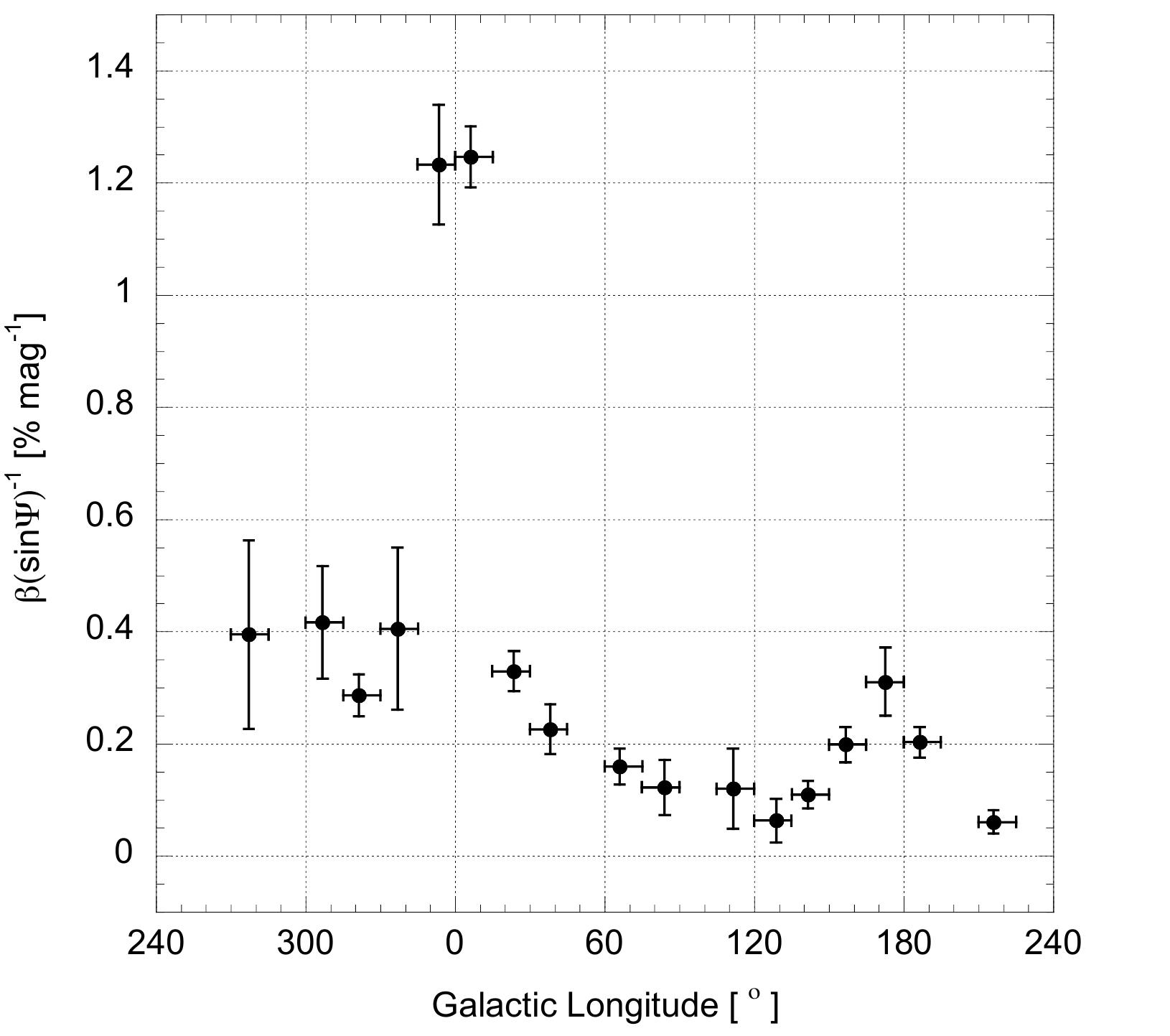}{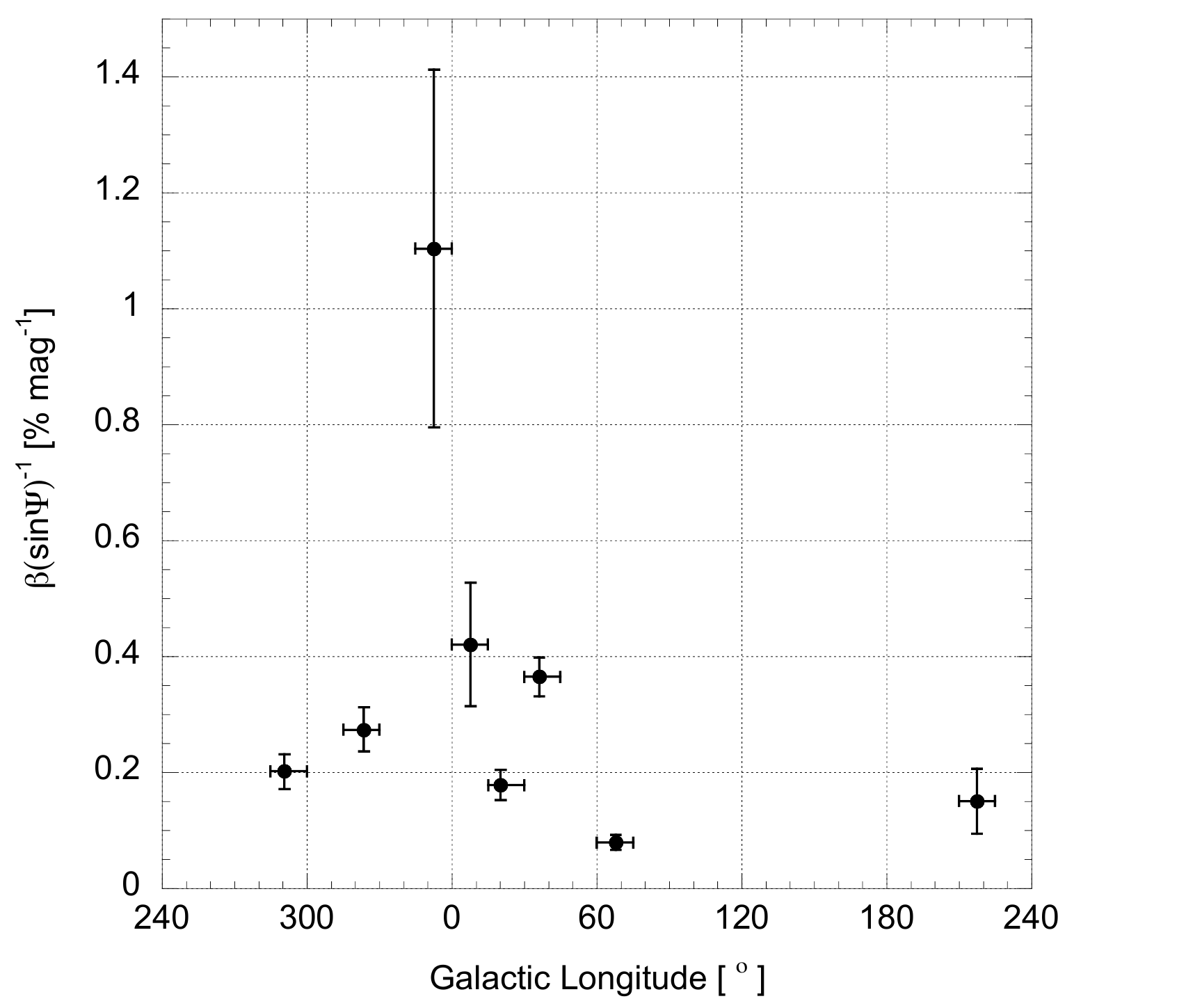}
	\caption{Plot of $\beta (sin\Psi)^{-1}$ as a function of Galactic longitude, for the range of $30^{\circ}<\textit{b}<41.8^{\circ}$ (left panel) and $41.8^{\circ}<\textit{b}<56.4^{\circ}$ (right panel).}
	\label{bsin_measured_only}
\end{figure*}

In contrast, as can be seen from Table \ref{summarytab}, significant spatial variability exists in the fractional polarization.  Figure \ref{bsin_measured_only} shows $\beta (sin\Psi)^{-1}$ for $30^{\circ}<\textit{b}<41.8^{\circ}$ and $41.8^{\circ}<\textit{b}<56.4^{\circ}$.  In the lower band two broad features at \textit{l}$\sim180$ and $330^{\circ}$ can be seen with two high-amplitude points superimposed on the latter.  In the $41.8^{\circ}<\textit{b}<56.4^{\circ}$ band only a single, lower amplitude, feature is clearly present, located at \textit{l}$\sim 300-60^{\circ}$. Due to the lack of data in some other regions, the presence of additional peaks cannot be ruled out in the middle band.

\subsubsection{Projection of the Galactic Magnetic Field} \label{projection_sec}

One global effect that could account for the observed variations in the observed fractional polarization is the projection of the local Galactic magnetic field.   As shown by e.g. \citet{axon1977,crutcher2003}, based on a sample of more distant stars than the present one, an ordered average magnetic field exists in the local spiral arm.  \citet{crutcher2003} derive a direction for this large-scale local field of \textit{l,b}$\sim$80.6,0.  Since the measured polarization should depend on the projection of the average field direction, via-a-vis the line of sight, we modeled the effect of such a projected Galactic field with $\beta=a+c\cdot sin(\textit{l}-80^{\circ})$ (Figure \ref{blow_both_mods}). Because we are here probing for the projection of a global magnetic field direction, we do not apply the $sin(\Psi)^{-1}$ factor in this fitting. 

\subsubsection{Radiation Field Variations} \label{radiation_sec}

Another effect that can cause variations in the fractional polarization can be taken from the observations of \citet{cashman2014}, who noted a spatial variation in Cloud 3 in Ldn\ 204, where the fractional polarization decreased with the projected distance from $\zeta$ Oph.  This effect is expected from radiation induced grain alignment, where the alignment should depend on the strength of the radiation field strength seen by the grain.

To test for variations due to radiatively driven grain alignment, we constructed a model to predict values of $\beta (sin\Psi)^{-1}$ for each of the sky bins and a given source distribution.  The model takes the form:
	
\begin{equation} \label{beta_r2}
\resizebox{.42 \textwidth}{!} 
{
$\beta (sin\Psi)^{-1}(l,b)\propto \frac{L_{*}}{(x_{*}-x_{LB})^{2}+(y_{*}-y_{LB})^{2}+(z_{*}-z_{LB})^{2}}$
}
\end{equation}
Where L$_{*}$ is the luminosity of the sources, \textit{x}$_{*}$ (etc.), refers to the source locations and \textit{x}$_{LB}$ (etc.), the center points for each bin on the bubble wall. As noted above (Section \ref{walldist}), $R_{LB}$, the radius of the Local Bubble cylinder for a given longitude range in the Galactic Plane, is assumed to be constant.   With the relatively large spatial binning of the polarimetry, the geometric center of the bin may not accurately reflect the center location of the data in the bins. To account for this, we used the weighted average midpoint for \textit{l}$_{LB}$ and \textit{b}$_{LB}$ as the center points of each area. Our model is illustrated in the center panel of Figure \ref{lb_diagram}.

Combining the sources, the model takes the final form:

\begin{equation} \label{modeq}
\beta (sin\Psi)^{-1}(l,b)=A+B\sum\limits_{i=1}^n \frac{L_{*,i}}{r^{2}_{i}}
\end{equation}
Where the free parameters $A$ and $B$ are used to linearly scale the radiative field to the alignment efficiency.  When fitting the model to the data, $L_{*}$, $d_{*}$, \textit{l}$_{*}$, \textit{b}$_{*}$ and $R_{LB}$ are held fixed while the A and B parameters are varied. 
    
The most likely source of a spatially varying radiation field at the Local Bubble wall distance is the non-uniform distribution of the nearby OB associations. \citet{dezeeuw1999} carried out a comprehensive census of the stellar contents of the OB associations within 1 kpc of the sun.  Here, we use their results for distances and locations (in \textit{l,b}) of the OB associations, along with the list of proposed stellar candidates for the OB associations.  For this analysis we treat each OB association as a single point source, integrating the stellar luminosities.  Cross checking the stellar identifiers from the \citet{dezeeuw1999} candidate list with the Tycho-2 Spectral Type Catalog \citep{wright2003}, we calculated the luminosity of each source using a table of physical stellar properties by spectral type provided in \cite{cox2000}.  Using the specific spectral type for our analysis provides more detail compared to using the \citet{dezeeuw1999} classifications of early-type and late-type (earlier and later than A0).

\begin{figure}[t]
	\centering
	\epsscale{1.2}
	\plotone{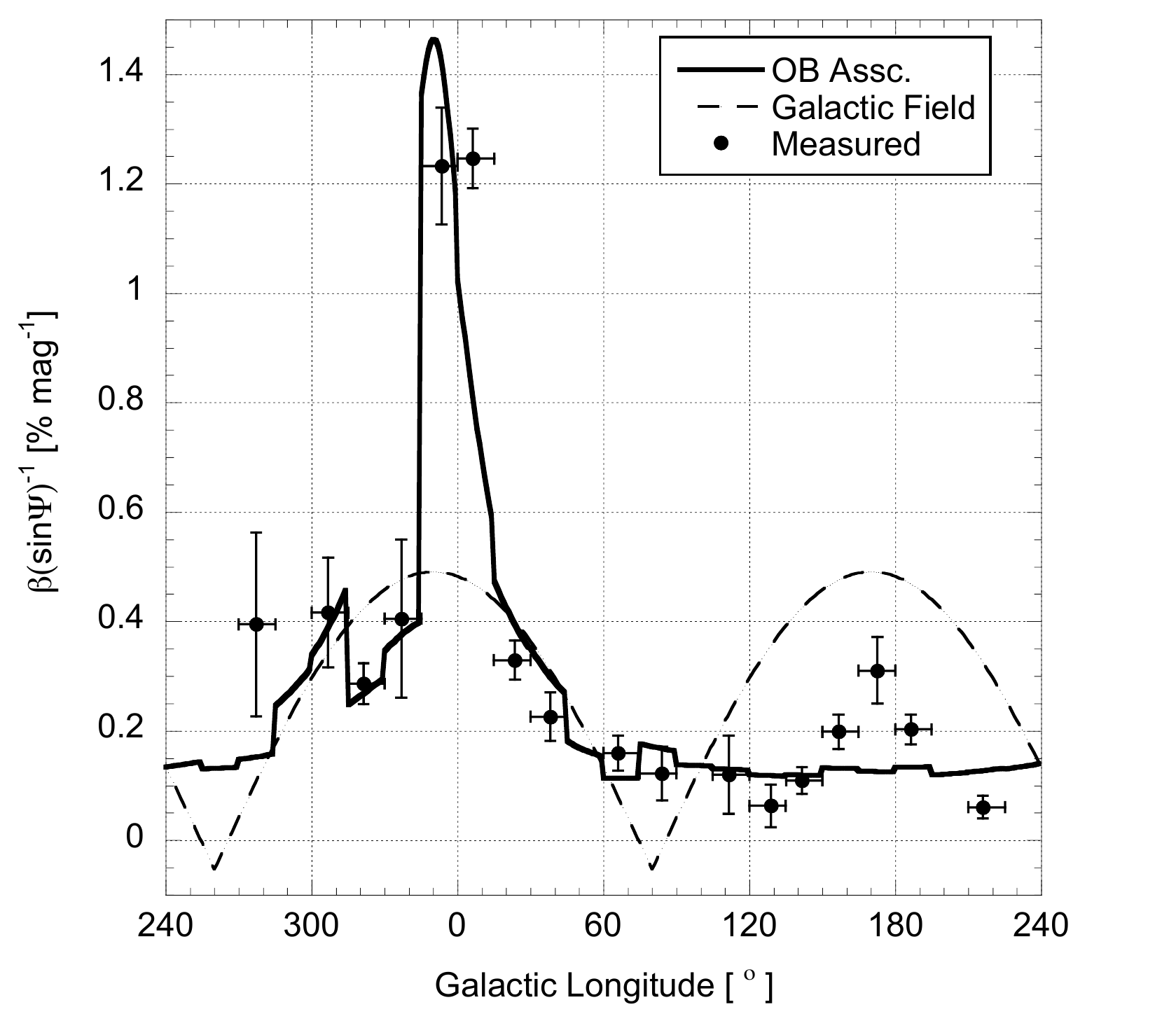}
	\caption{Plot of $\beta$ as a function of Galactic longitude, for the range of $30^{\circ}<\textit{b}<41.8^{\circ}$. Overlaid is the best fit for the model based the projection of a Galactic magnetic field (dashed line, Section \ref{projection_sec}), which yields a best fit of $a=-0.00218\pm 0.15880$ and $c=0.436\pm0.207$, and the best fit for the model based on radiatively driven alignment (full drawn, Section \ref{radiation_sec}), which yields a best fit of $A=0.00547\pm0.01028$ and $B=0.00469\pm0.00021$.}
	\label{blow_both_mods}
\end{figure}

\begin{figure}[t]
	\centering
	\epsscale{1.2}
	\plotone{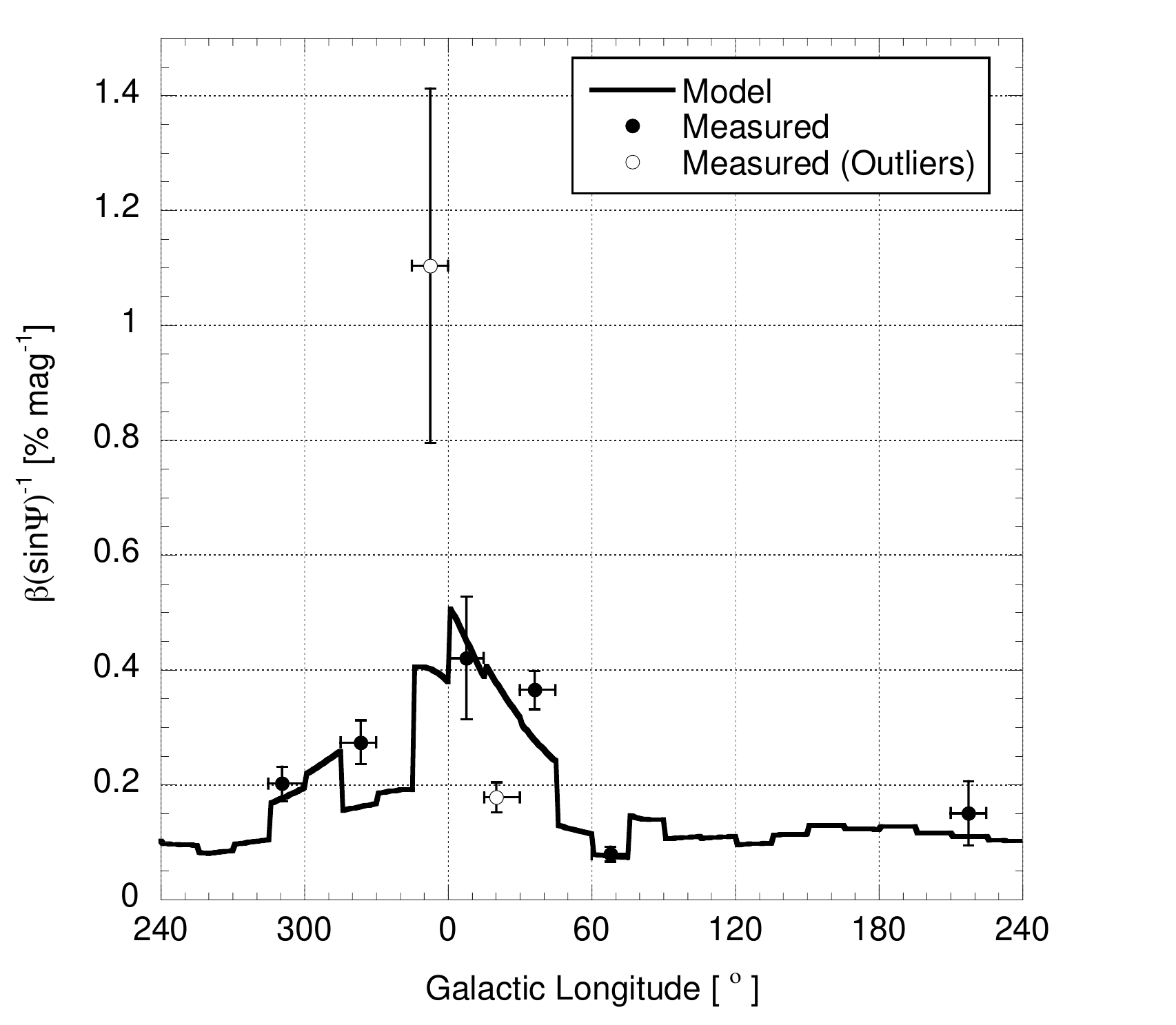}
	\caption{Plot of the $\beta (sin\Psi)^{-1}$ as a function of Galactic longitude with the OB association model function overlaid for the range of $41.8^{\circ}<\textit{b}<56.4^{\circ}$. The best fit yields $A=0.00547\pm0.01028$ and $B=0.00469\pm0.00021$.}
	\label{bmidband}
\end{figure}

The derived characteristics for each association are summarized in Table \ref{obsumtab}. The Cas-Tau association, while included in \citet{dezeeuw1999}, was not included in our OB-association model due to its very wide extent over the sky.  The large angle distribution means it cannot be well approximated with a localized source and would be equivalent to a zero-point offset. The best fit to the model using only the OB associations as the sources is shown as the full-drawn line in Figure \ref{blow_both_mods} and Figure \ref{bmidband}

\begin{figure}[t]
	\begin{deluxetable}{lcccc}
		\tablecaption{Summary of OB association characteristics. \label{obsumtab}}
		\tablehead{\colhead{Name} & \colhead{$d_{OB}$\tablenotemark{$\dag$}} & \colhead{\textit{l$_{min}$} \ \textendash \ \textit{l$_{max}$}\tablenotemark{$\ddag$}} & \colhead{\textit{b$_{min}$} \ \textendash \ \textit{b$_{max}$}\tablenotemark{$\ddag$}} & \colhead{$log(L_{tot})$} \\
			\colhead{} & \colhead{[pc]} & \colhead{[$^{\circ}$]} & \colhead{[$^{\circ}$]} & \colhead{[$log(L_{\sun})$]}}
		\startdata
		US      & 145 & 343 \ \textendash \ 360 & 10 \ \textendash \ 30 & 5.563 \\
		UCL     & 140 & 312 \ \textendash \ 343 & 0 \ \textendash \ 25  & 5.270 \\
		LCC     & 118 & 285 \ \textendash \ 312 & -10 \ \textendash \ 21   & 4.908 \\
		VOB2 & 410 & 255 \ \textendash \ 270 & -15 \ \textendash \ -2   & 5.716 \\
		T10 & 366 & 255 \ \textendash \ 270 & -2 \ \textendash \ 4    & 4.481 \\
		C121 & 592 & 228 \ \textendash \ 244 & -15 \ \textendash \ -2   & 6.132 \\
		POB2 & 318 & 145 \ \textendash \ 170 & -27 \ \textendash \ -13   & 5.514 \\
		$\alpha$P & 177 & 140 \ \textendash \ 165 & -17 \ \textendash \ 3   & 4.303 \\
		LOB1 & 368 & 90 \ \textendash \ 110  & -25 \ \textendash \ -5   & 5.214 \\
		COB2 & 615 & 96 \ \textendash \ 108  & -5 \ \textendash \ 14   & 6.102 \\
		COB6 & 270 & 100 \ \textendash \ 110 & -2 \ \textendash \ 2    & 4.641 \\
		\enddata
		\vspace{3mm}
		\tablenotetext{$\dag$}{From Table 2 in \citet{dezeeuw1999}}
		\tablenotetext{$\ddag$}{From Table A1 in \citet{dezeeuw1999}}
		\tablecomments{Full names \ of OB associations\ in column \ one are: \\US: Upper Scorpius, UCL: Upper Centaurus Lupus, LCC: \\Lower Centaurus Crux, VOB2: Vela OB2, T10: Trumpler \\10, C121: \ Collinder 121, \ POB2: \ Perseus\ OB2, \ $\alpha$P: \ $\alpha$\\ Persei, LOB1: \ Lacerta OB1, COB2: \ Cepheus \ OB2 and \\COB6: Cepheus OB6}
	\end{deluxetable}
\end{figure}

While the general population of field stars in the Galactic neighborhood does not show a strong spatial variability, it is \textbf{nonetheless} important to estimate the contributions from stars not included in the \citet{dezeeuw1999} survey.  To gauge the effects from such ``field stars" we constructed a catalog using the sources from the Michigan spectral classification survey (MSS) of the Henry Draper (HD) catalog \citep{michiganV1,michiganV2,michiganV3,michiganV4,michiganV5}.  This choice is based on the uniformity of the HD catalog and the high quality of the spectral classification. Since the published MSS primarily covers the Southern sky, we filled in areas not covered by MSS with sources from \citet{wright2003} to accrue an all-sky catalog. Due to the higher uncertainty in the classifications from \citet{wright2003}, we only included sources with $m_V<9$, as we expect the brighter sources to have more reliable spectral classifications and generally be more important for the local radiation field (this magnitude limit is similar to that of the MSS).  We used Gaia \citep{gaiaDR1,gaiaDR2} and Hipparcos \citep{hip1997} parallax based distances and included only those stars with $d/\sigma_d>2.5$.  A total of 126,105 stars met these criteria.  We note that the bulk of the members of the Cas-Tau association, excluded above, were included in this sample.

\begin{figure*}[t]
	\centering
	\includegraphics[width=1.0\linewidth]{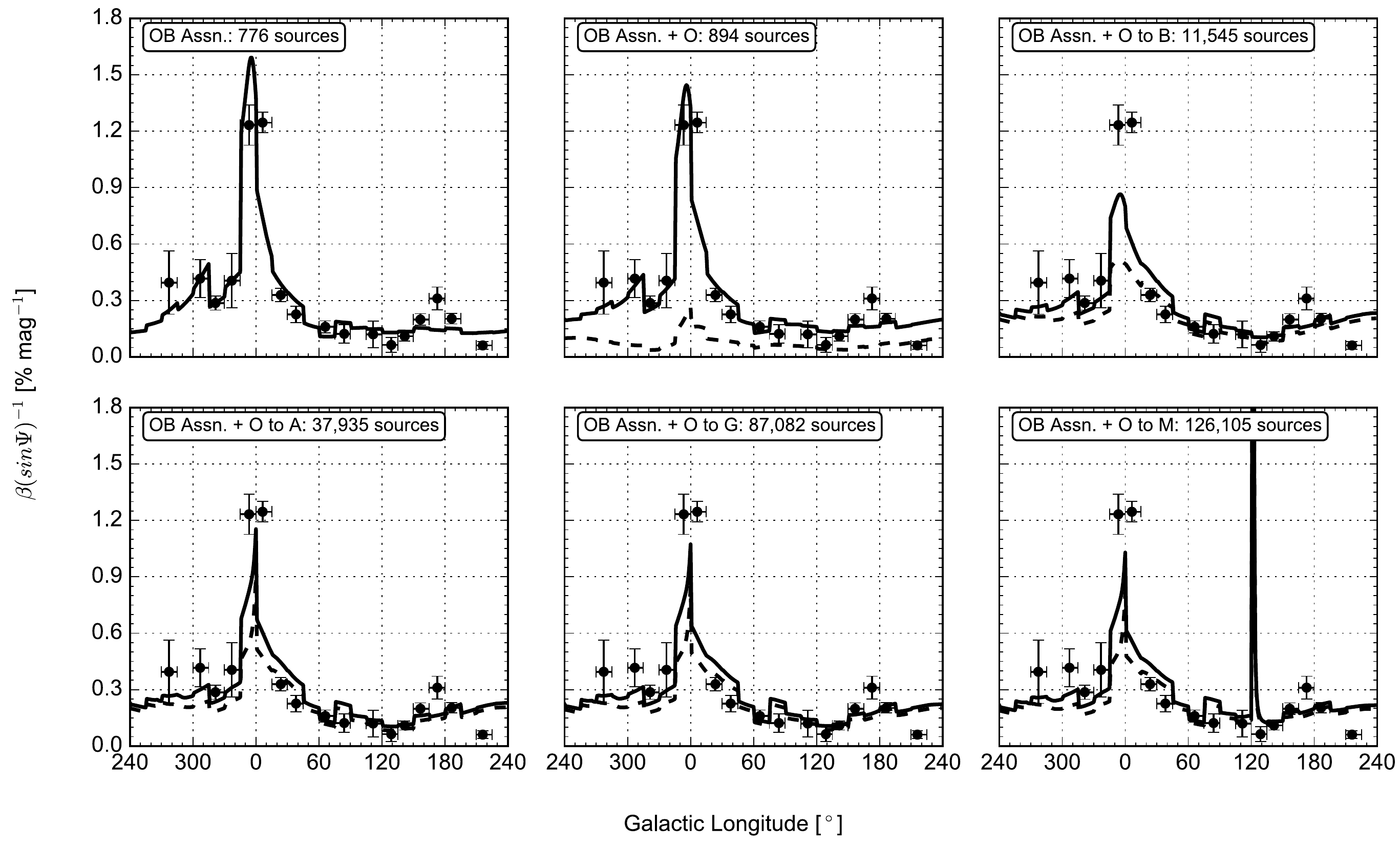}
	\caption{Plot of $\beta (sin\Psi)^{-1}$ as a function of Galactic longitude for the model of point sources in the field for the range of $30^{\circ}<\textit{b}<41.8^{\circ}$. In this figure, the points are the measured data, the solid line is the best fit to the model, and the dashed line is the model minus the contribution from the sources in the OB associations. The temperature class of stars and how many stars are included within a given sample are listed in the top left corner of each panel.}
	\label{pointmodlow}
\end{figure*}

\begin{figure*}[t]
	\centering
	\includegraphics[width=1.0\linewidth]{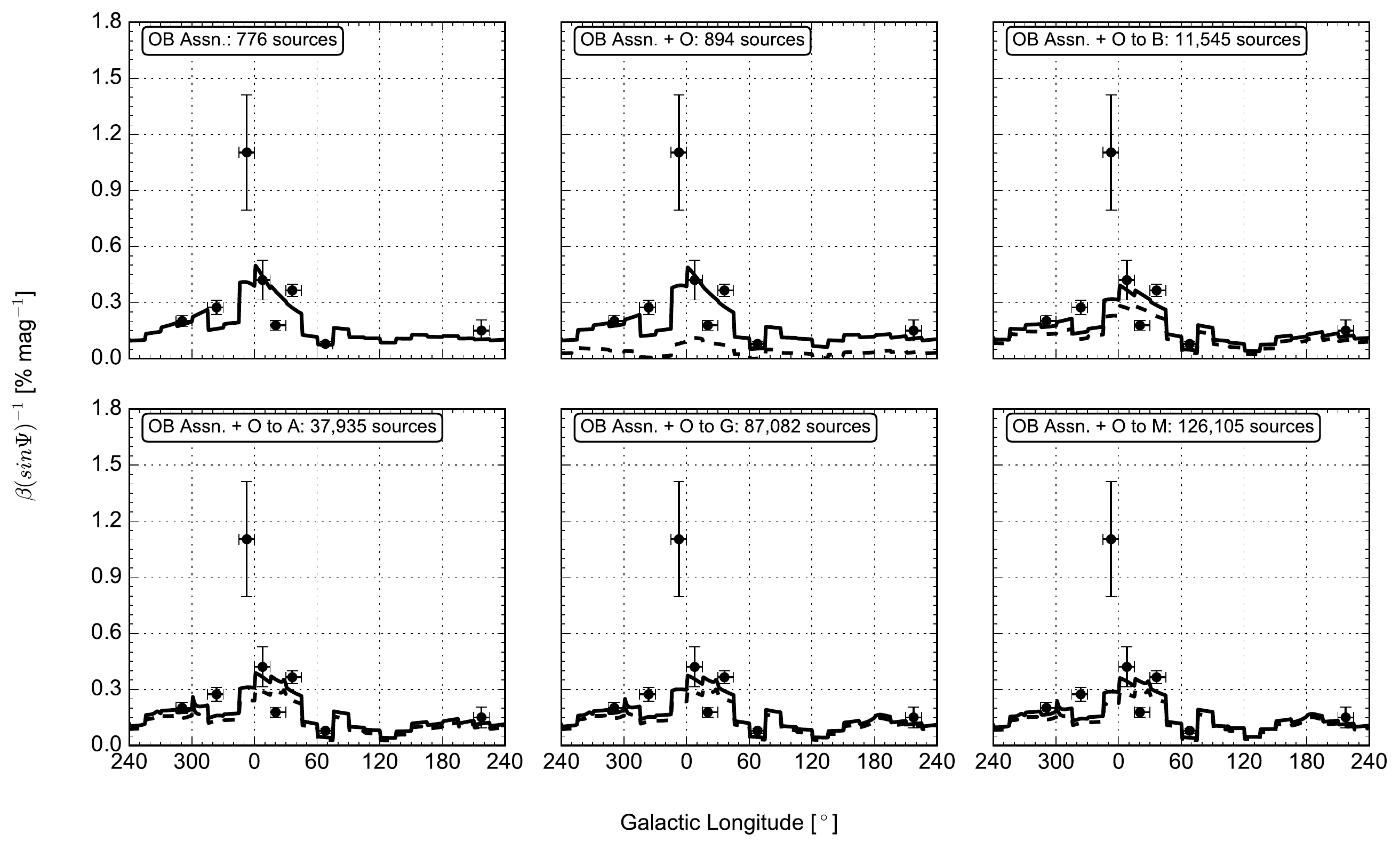}
	\caption{Plot of $\beta (sin\Psi)^{-1}$ as a function of Galactic longitude for the model of point sources in the field for the range of $41.8^{\circ}<\textit{b}<56.4^{\circ}$. In this figure, the points are the measured data, the solid line is the best fit to the model, and the dashed line is the model minus the contribution from the sources in the OB associations. The temperature class of stars and how many stars are included within a given sample are listed in the top left corner of each panel.}
	\label{pointmodmid}
\end{figure*}

One complication that arises in such a field star sample is that, because of the $1/r^2$ dependence of the radiation intensity, combined with the uncertainties in spectral classification parameters, and in the distances to the stars and the Local Bubble wall, there might be sources that could, given distance and luminosity uncertainties, artificially dominate specific regions of the modeled radiation field at the Local Bubble wall.  In particular, we note four stars where this might be the case. The MSS classification of HD 127493, as a B0, is likely incomplete and should likely be a sub-dwarf rather than luminosity class V \citep{krtivcka2016}. $\zeta$ Puppis, HD 46223 and HD 206267 seem to have overestimated luminosities based on the average spectral class vs. luminosity relations.  Dedicated studies of these stars provide luminosities of $\sim800,000L_{\sun}$ \citep{repolust2004,fullerton2006,najarro2011}, $\sim450,000L_{\sun}$ \citep{wang2008, martins2012} and $\sim400,000L_{\sun}$ \citep{waldron2007}, respectively. In our field star model we therefore use these updated parameters.  The best-fit models using field stars as the radiation field source are shown in Figure \ref{pointmodlow} and Figure \ref{pointmodmid}.

As the spike at l$\sim120^\circ$ in panel six of Figure \ref{pointmodlow} shows, additional inaccuracies or issues may still be included in the model (this particular peak is due to the K0~Ib star HD~115337 at d=223$\pm$3 pc).  It is, of course, also possible that other field stars have catalog parameters which would under-estimate their actual contributions to the radiation field at the Local Bubble wall, which could explain some poorly fit minor features in the fractional polarization, such as at \textit{l}$\approx$180 in the lower band (Figure \ref{blow_both_mods}).  However, given the size of the sample, such deviations are unlikely to dominate the fitting.

\subsection{Position Angle Dispersions} \label{sec:pol_ang_disp}
\begin{figure*}[t]
	\centering
	\includegraphics[width=1.0\linewidth]{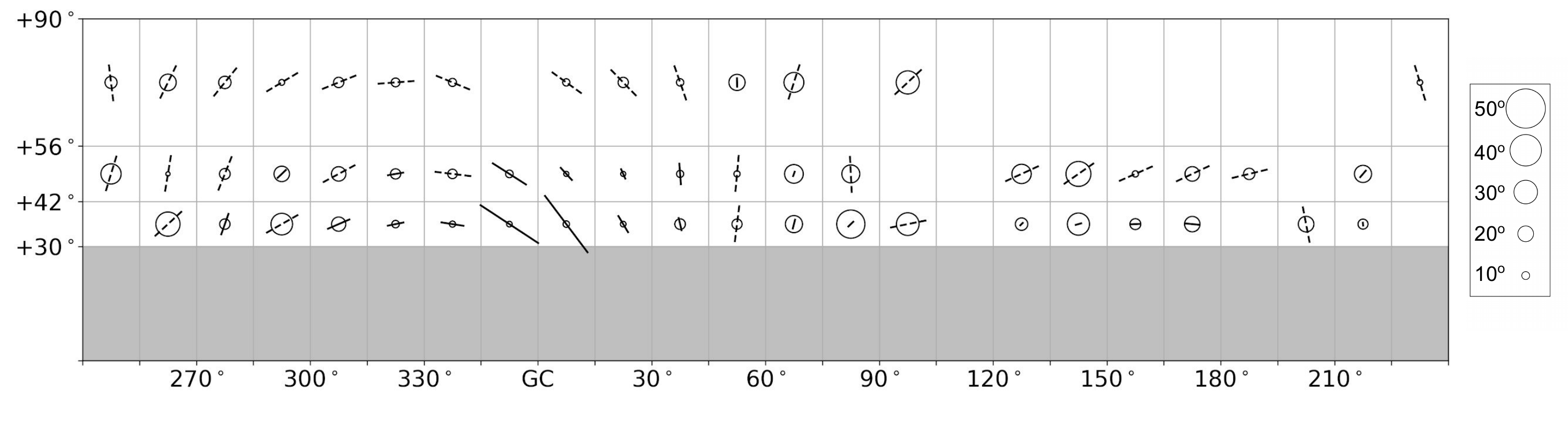}
	\caption{Plot of the polarization direction, fractional polarization, as parametrized by $\beta$ (indicated by the direction and length of the lines), and the polarization angle dispersion (indicated by the size of the circle centered at each line) over the survey area. Note that for the regions where lines are plotted with dashed line, too few values of the visual extinction are available, and therefore no fit of the fractional polarization was possible.  The length of these bars have therefore been set to a fixed, arbitrary, length. The areas in solid gray correspond to regions not covered in the \citet{berdyugin2014} sample.}
	\label{PAplot}
\end{figure*}

\begin{figure}[t]
	\epsscale{1.2}
	\plotone{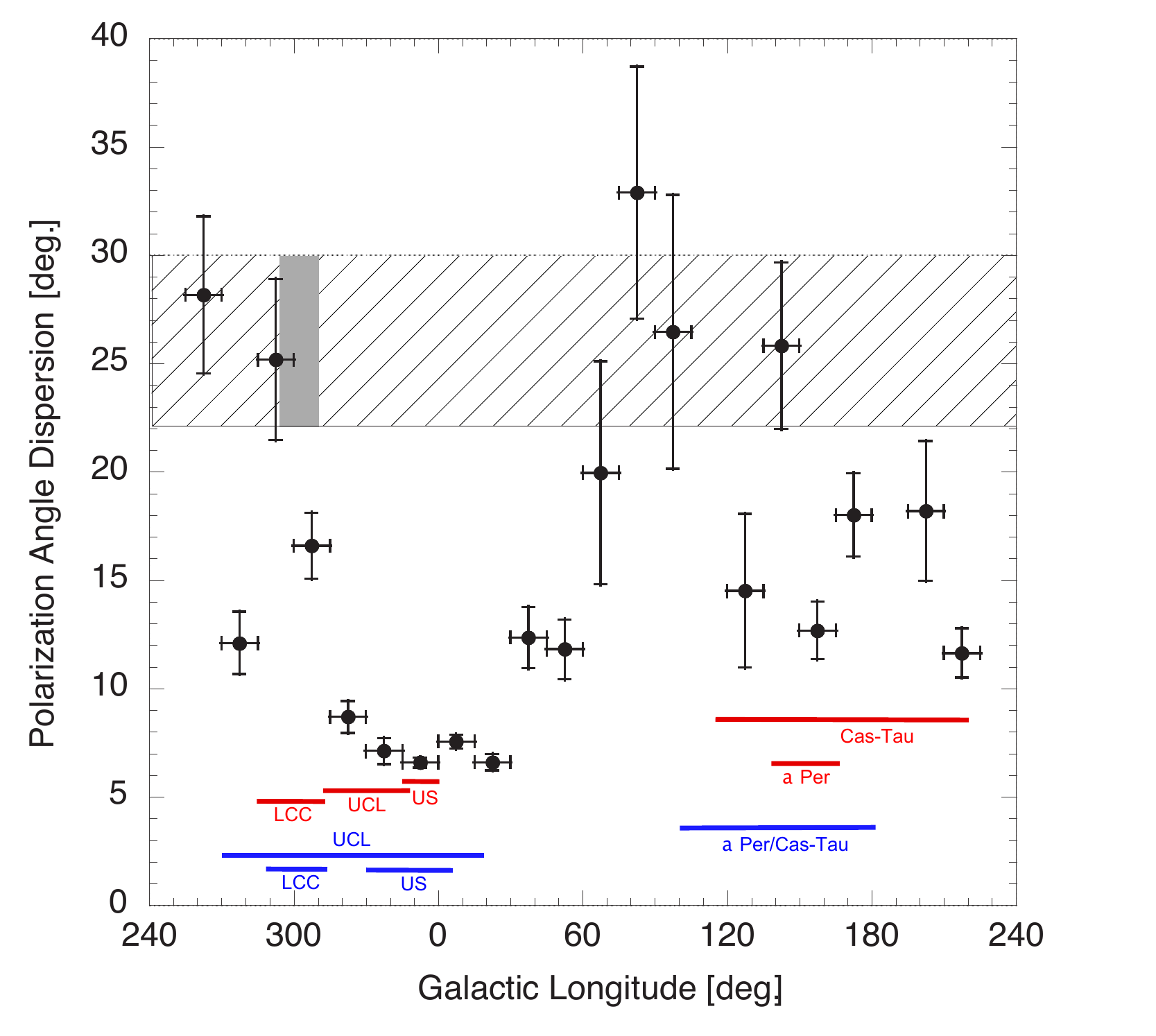}
	\caption{Plot of polarization angle dispersion as a function of Galactic longitude for the range of $30^{\circ}<\textit{b}<41.8^{\circ}$ (filled circles). The dispersion found by \citet{bga2006} towards \textit{l,b}$\approx300,0^\circ$ is shown as the gray rectangle over the approximate extent of the Southern Coalsack, with the hatched region extended over the full longitude range for reference. The approximate extents of the OB associations from \citet{dezeeuw1999}, and the HI super bubbles from \citet{degeus1992} and \citet{bhatt2000}, within 200 pc are shown as red and blue bars, respectively.}
	\label{glonvspa}
\end{figure}

As discussed above, the Davis-Chandrasekhar-Fermi method \citep{davis1951b,chandrasekhar1953} can be used to estimate the magnetic field strength in a plasma by measuring local variations in the orientation of the field lines.  The observable from polarimetry data is the dispersion of the position angles within a limited spatial region. An important observational caveat is the mixing of unrelated regions (turbulent cells) along the line of sight and across the plane of the sky.  We can use the measured line widths of absorption line data to constrain the former (Section \ref{turb}). We will assume that the spatial binning used above is limited enough to not include large-scale field variations in a given sample. The measured dispersions are listed in Table \ref{summarytab}, and plotted in Figures \ref{PAplot} and \ref{glonvspa}. 

\section{results}\label{sec:results}

\subsection{Polarization Projection}
The best fit to the model representing the effects of a projected Galactic magnetic field is shown in Figure \ref{blow_both_mods} as the dashed curve.  The poor resulting fit indicates that a projected Galactic field alone is unlikely to be able to account for the variations in the observed polarization fraction, as neither the amplitude or the width of the observed variations match this model.

\subsection{Grain Alignment}\label{subsection:results,grain_alignment}
\subsubsection{Radiative Alignment due to Local OB Associations}  \label{radtoOB}
\begin{figure}[t]
	\centering
	\epsscale{1.2}
	\plotone{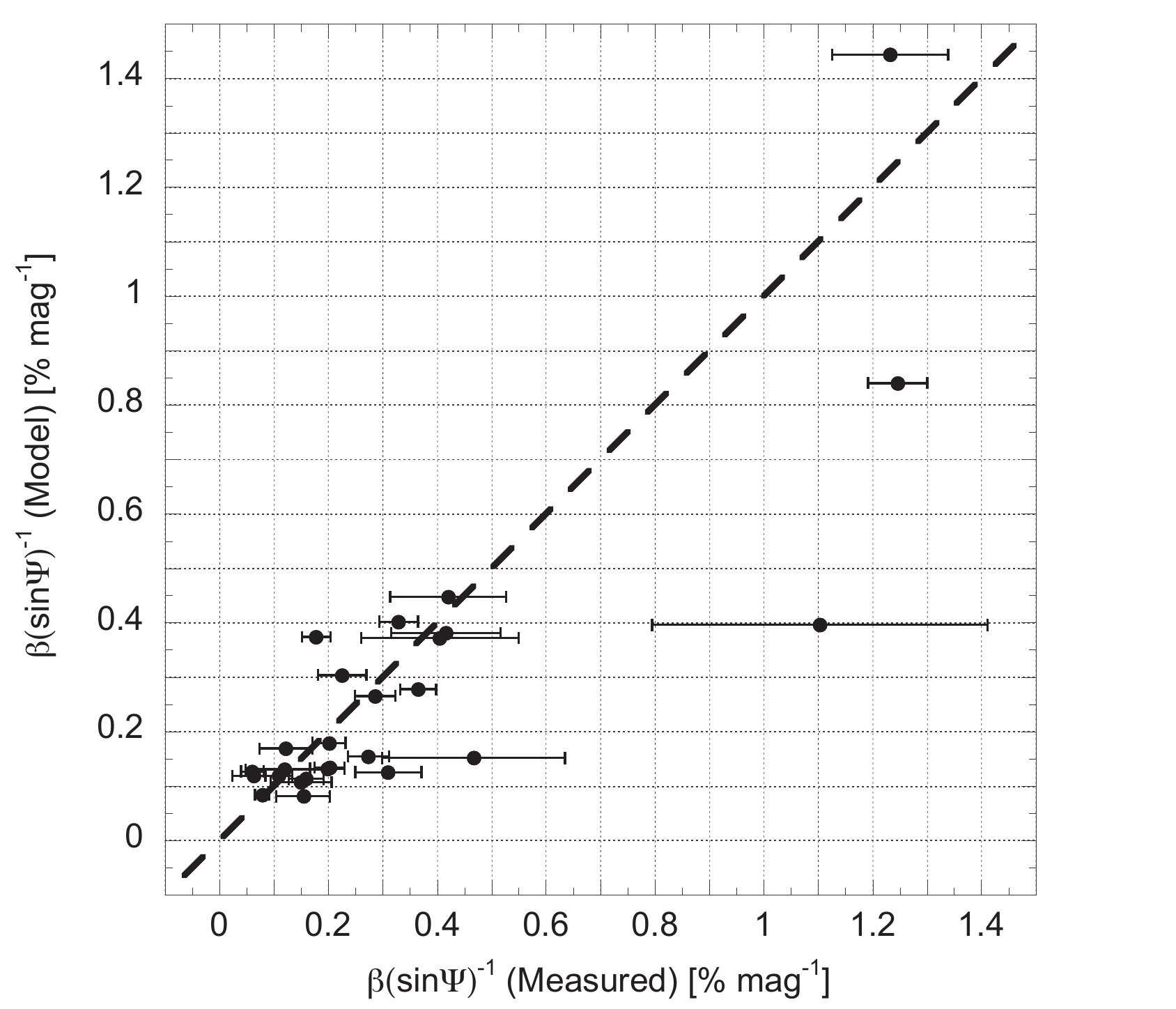}
	\caption{Plot of the model predictions for radiatively driven grain alignment based on the local OB-associations, versus measured $\beta (sin\Psi)^{-1}$ values, including all areas in Table \ref{summarytab}.  The dashed line shown is the line for a 1-to-1 relationship. The best fit yields $A=0.00547\pm0.01028$ and $B=0.00469\pm0.00021$.}
	\label{bcompareolsweight}
\end{figure}

The fits resulting from a model of radiatively driven grain alignment, using the local OB-associations as sources, are shown in Table \ref{modelsumtab} and over-plotted in Figures \ref{blow_both_mods}, \ref{bmidband} (full drawn lines), and \ref{bcompareolsweight}. As discussed above, for these fits, we have treated each OB association as a point source with a flux given by the summed stellar luminosities. The discontinuities in Figures \ref{blow_both_mods} and \ref{bmidband} are due to the estimation of the Local Bubble wall distance as a constant value for each region, creating a discretized model function rather than a continuous one.  While our model is fit only to the mid-points of each slice, in these figures the function is plotted on a finer grid, giving rise to the slope in the model plots within bins.  Good over-all fits are achieved to the data.

For the lower band sample ($30^{\circ}<\textit{b}<41.8^{\circ}$; Figure \ref{blow_both_mods}), we find a reduced $\chi^2$ value of $\chi^2_{15}=7.0$.  For the mid-latitude-band ($41.8^{\circ}<\textit{b}<56.4^{\circ}$; Figure \ref{bmidband}) the projection-corrected data yields $\chi^2_{15}=12.2$.  Comparing fits of the projection-corrected value $\beta (sin\Psi)^{-1}$ with those of $\beta$ alone, we find, for the lower band, a reduced $\chi^2$ \textbf{value} for the latter of $\chi^2_{15}=15.7$.  Hence the inclusion of the $(sin\Psi)^{-1}$ factor provides a significantly better fit to the data.  

The formally somewhat large values of $\chi^2$=7.0 and 12.2, even for the projection-corrected data, are likely due to the inherent simplifications in our model assumptions, and possible under-estimations of some individual errors. Figure \ref{blow_both_mods} shows that two data points, in particular, contribute the most to the calculated $\chi^2_{\nu}$ value; the areas at \textit{l}$\approx$10$^\circ$ (Area 1) and 270$^{\circ}$ (Area 55). The model mismatch near the Galactic Center direction can most likely be attributed to the large spatial binning of the data.  If we use a longitude value at the low-longitude edge of the bin (\textit{l}$=0^{\circ}$), the model fit becomes within 1$\sigma$ of the data.   For the point at $\sim270^{\circ}$, the somewhat poor model fit may be the result of an incorrect bubble wall distance caused by the model assumptions; that the wall distance is well modeled as a cylinder, with a fixed radius for each longitude.  For Area 55, the maps by \citet{lallement2003} show that the Local Bubble wall radius for \textit{b}$>30^\circ$ is significantly less than what is observed along the Galactic plane. No other region shows as significant a variation in $R_{LB}$ with \textit{b}, and our assumption should therefore hold for the rest of the bins.

The final region of significant deviations for $30^{\circ}<\textit{b}<41.8^{\circ}$ (Figure \ref{blow_both_mods}) is $l\sim130-200^{\circ}$.  These are clearly systematic in nature, indicating that an element may be missing from our model, such as a high-luminosity star (luminosity class I-III) misclassified as a Main Sequence (lum. class V) source, located close to the Bubble wall.

For the middle latitude band there are two data points that are not well modeled (Figure \ref{bmidband}; open circles). The value for \textit{l}$=345^{\circ}-360^{\circ}$ is, however, quite uncertain.

\subsubsection{Radiative Alignment due to Local Field Stars}  
\begin{figure}[t]
	\centering
	\epsscale{1.2}
	\plotone{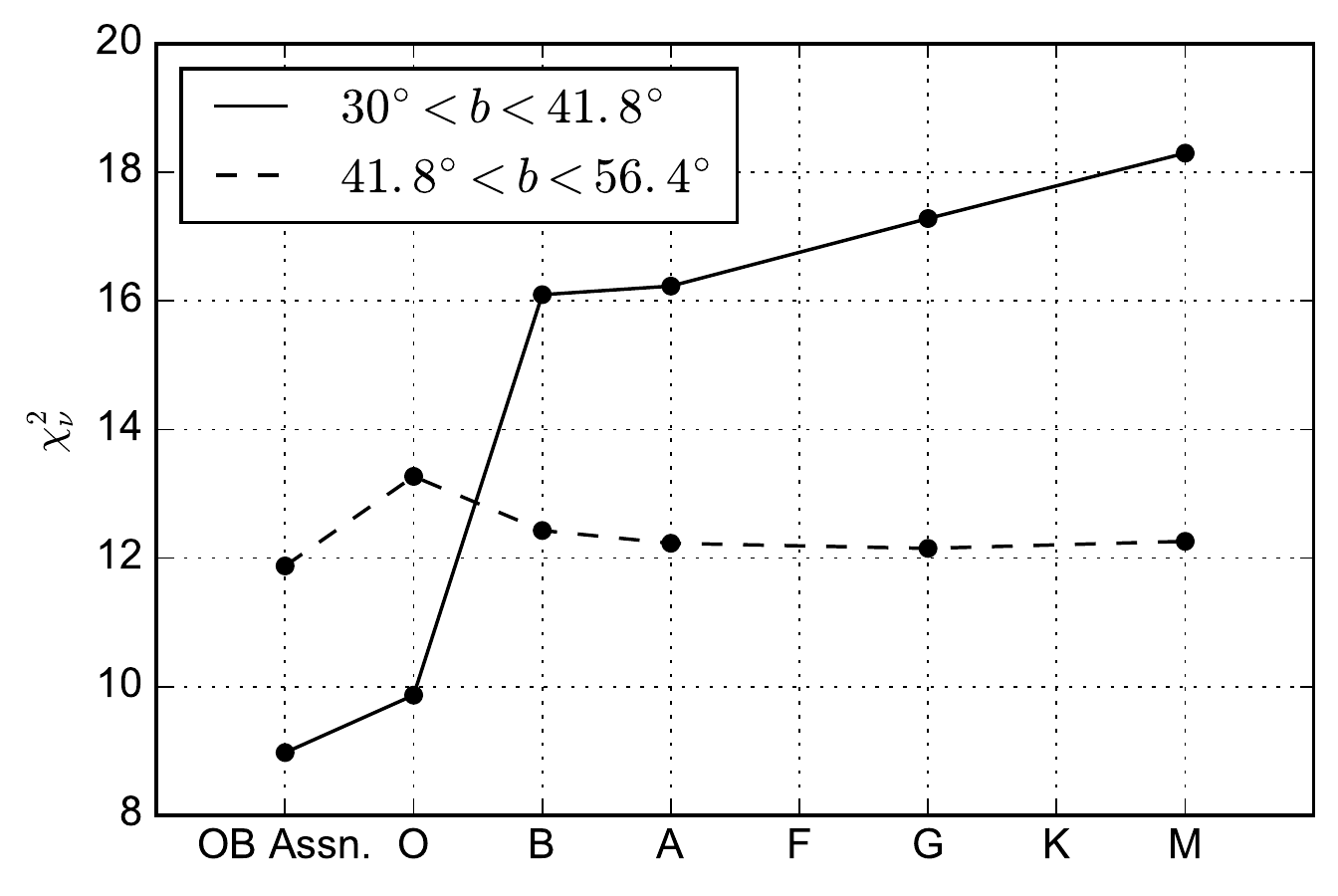}
	\caption{Reduced $\chi^2$ values for the model fits in Figure \ref{pointmodlow} (solid line) and Figure \ref{pointmodmid} (dashed line).}
	\label{chisqumodels}
\end{figure}

Using the field star sample as the sources of the radiation and adding in consecutive spectral classes from our field star sample to the OB-association member stars produce the fits shown in Figures \ref{pointmodlow} and \ref{pointmodmid}. The resulting reduced $\chi^2$ values are shown in Figure \ref{chisqumodels}.  

The total number of stars included in each model fit are indicated in the sub-plots.  In these fits, the OB-association member stars were treated as individual point sources.

As would be expected, the fits incorporating only O and B stars are very similar to the fits for the OB-association above, and match the data very well.  As shown by these fits and the $\chi^2$ values for the lower latitude band, the models with only blue sources (i.e. the O and B stars) produce better fits to the fractional polarization (reduced $\chi^2\approx 9$) than more inclusive models (reduced $\chi^2\approx 18$, with all spectral classes), while all models \textbf{produce} similar results for the middle latitude band.

\begin{figure}[t]
\begin{deluxetable}{lcccc}
	\tablecaption{Summary of model results compared to measured data. \label{modelsumtab}}
	\tablehead{\colhead{Area} & \colhead{$R_{LB}$} & \colhead{$\beta (sin\Psi)^{-1}$} & \colhead{$\textit{l}_{mid}$,  $\textit{b}_{mid}$}& \colhead{$\beta (sin\Psi)^{-1}$ (Model)}\\
		\colhead{}  & \colhead{[pc]} & \colhead{[$\%mag^{-1}$]} & \colhead{[$^{\circ}$]} & \colhead{[$\%mag^{-1}$]}}
	\startdata
1  & 100 & $1.25\pm0.05$ & 6.3, 35.20    & 0.84 \\
2  & 100 & $0.42\pm0.11$ & 7.8, 46.82   & 0.45 \\
4  & 75  & $0.33\pm0.04$ & 23.7, 37.07  & 0.40  \\
5  & 75  & $0.18\pm0.03$ & 20.3, 48.28  & 0.37 \\
7  & 80  & $0.23\pm0.04$ & 38.2, 35.48  & 0.30  \\
8  & 80  & $0.37\pm0.03$ & 36.3, 48.46  & 0.28 \\
12 & 150 & $0.15\pm0.05$\tablenotemark{$\dag$} & 50.0, 63.09    & 0.08  \\
13 & 210 & $0.16\pm0.03$ & 66.0, 34.08    & 0.11 \\
14 & 210 & $0.08\pm0.01$ & 68.0, 50.94    & 0.08 \\
16 & 95  & $0.12\pm0.05$ & 84.1, 36.12  & 0.17 \\
22 & 160 & $0.12\pm0.07$ & 111.6, 34.09 & 0.13 \\
25 & 180 & $0.06\pm0.04$ & 128.9, 36.41 & 0.12 \\
28 & 175 & $0.11\pm0.02$ & 141.7, 35.4  & 0.12 \\
31 & 130 & $0.20\pm0.03$  & 156.7, 36.78 & 0.13 \\
34 & 145 & $0.31\pm0.06$ & 172.7, 35.93 & 0.13 \\
37 & 130 & $0.20\pm0.03$  & 186.4, 34.43 & 0.13 \\
43 & 160 & $0.06\pm0.02$ & 216.0, 36.27   & 0.13 \\
44 & 160 & $0.15\pm0.06$\tablenotemark{$\dag$} & 217.5, 49.61 & 0.11  \\
55 & 200 & $0.47\pm0.17$ & 277.2, 34.76 & 0.15 \\
59 & 140 & $0.20\pm0.03$  & 290.8, 47.62 & 0.18 \\
61 & 130 & $0.42\pm0.10$  & 306.8, 36.63 & 0.38 \\
64 & 180 & $0.29\pm0.04$ & 321.4, 35.46 & 0.27 \\
65 & 180 & $0.27\pm0.04$ & 323.7, 46.82 & 0.15 \\
67 & 170 & $0.41\pm0.14$ & 337.0, 35.68   & 0.37 \\
70 & 120 & $1.23\pm0.11$ & 353.5, 34.72 & 1.44 \\
71 & 120 & $1.10\pm0.31$  & 352.7, 47.18 & 0.40 
	\enddata
	\vspace{3mm}
	\tablenotetext{$\dag$}{Regions where $\Psi\leq13^\circ$, so there is no correction due to the \\projection effect.}
	\tablecomments{The second column is the measured $\beta (sin\Psi)^{-1}$ data \\and the subsequent  columns are the \textit{l, b} inputs and  $\beta (sin\Psi)^{-1}$ \\results from the model.}
\end{deluxetable}
\end{figure}

\subsection{Position Angle Dispersions}\label{subsection:results,dispersion}

Figures \ref{PAplot} and and \ref{glonvspa} show significant, coherent, variability in the position angle dispersion over the sky, with a notable systematic low dispersion towards the inner Galaxy.  For some other regions the measured position angle dispersions are similar to that found for the Local Bubble wall in the direction of the Southern Coalsack (\textit{l,b}=300,0), $\Delta\theta=26\pm4^\circ$ \citep[indicated by the gray rectangle in Figure \ref{glonvspa}]{bga2006}.

The analysis of \citet{bga2006} is limited to the area covering the projected extent of the Southern Coalsack cloud.  As noted above, the gas density and velocity dispersion for the material were derived from ultraviolet observations of the fine structure lines of neutral carbon \citep{jenkins1979} towards the star $\mu^2$ Cru ($\textit{l,b,d}\approx303^\circ,6^\circ,112 pc$) directly constraining the turbulent velocity of the relevant gas.  The C I excitation yielded a thermal gas pressure of $\sim$10,000-20,000 K cm$^{-3}$, but with large uncertainties.  Utilizing far ultraviolet observations of the H$_2$ excitation in bubble wall gas \citep{lehner2003}, to constrain the gas temperature \citet{bga2006} estimated a gas density of $n=50^{+25}_{-5}\ cm^{-3}$ and a magnetic field of $B_\perp=8^{+5}_{-3}\mu G$.  This field strength is equivalent to a magnetic pressure in the Local Bubble wall of $P/k=18,000^{+23,000}_{-10,000}\ K\ cm^{-3}$ which is close to both the gas pressure in the bubble wall, from the C I excitation, and the estimated pressure in the Local bubble from X-ray and EUV data \citep{bowyer1995,snowden1998}.  Given the more complete data set for the Coalsack analysis, we discuss the current polarization angle dispersions in the context of these earlier results.

To quantify the variations in the distribution of polarization angle dispersions, we fitted one- and two-component Gaussians to the  measurements.  While the distribution of angle dispersions is likely more complicated than a two-value bifurcation, we use this as a simple test for the reality of the variability.  We found a best one-component fit of $\langle\Delta\theta\rangle=12.4\pm2.0$ with a $\chi^2=77.3$, while the two component fit results in $\langle\Delta\theta\rangle_1=7.30\pm0.72$ and $\langle\Delta\theta\rangle_2=16.2\pm2.9$ with a $\chi^2=27.1$, yielding \textbf{ratio} of dispersions of 2.2.  \textbf{Hence, the data statistically support a multi-valued  dispersion distribution.}  The ``small" value can be seen to be close to the coherent structure towards the inner Galaxy in Figure \ref{glonvspa}.  The ``large" value may be an under-estimate of the maximum dispersions in the data if a single ``large dispersion" is in reality a graduated - or multi-valued - set.  For instance, several regions have $\Delta\theta$ close to the value seen by \citet{bga2006} of $\langle\Delta\theta\rangle_1\approx26^{\circ}$.  Using the \citet{bga2006} value, we find a ratio between a ``large" and ``small" angle dispersions of $\sim3.6$.  \textbf{Such a difference in angle dispersions could imply a significant variation in the magnetic field strength over the local bubble wall, with significant dynamical consequences for the medium.}

We can use the extent of OB associations within 200pc of the Sun to explore the origin of these angle dispersion variations.  In Figure \ref{glonvspa} we have indicated both the extent of the OB associations themselves (red bars) from \citet{dezeeuw1999}, as well as the extent of the associated super bubbles (blue bars) as probed by \citet{degeus1992} in H I (Sco-Cen) and \citet{bhatt2000} in CO (Cas-Tau/$\alpha$ Per).  A general correspondence between the extent of the near-by OB associations and super bubbles, and the low disperion regions is apparent, especially towards the inner Galaxy and Sco-Cen.  Because the centers of the associations/bubbles are generally below the latitude probed by the \citet{berdyugin2014} sample, and because of structures of the intervening ISM, we would not expect a perfect correspondence.   We explore the correlation as a working hypothesis below.

\begin{figure}[t]
	\centering
	\epsscale{1.2}
	\plotone{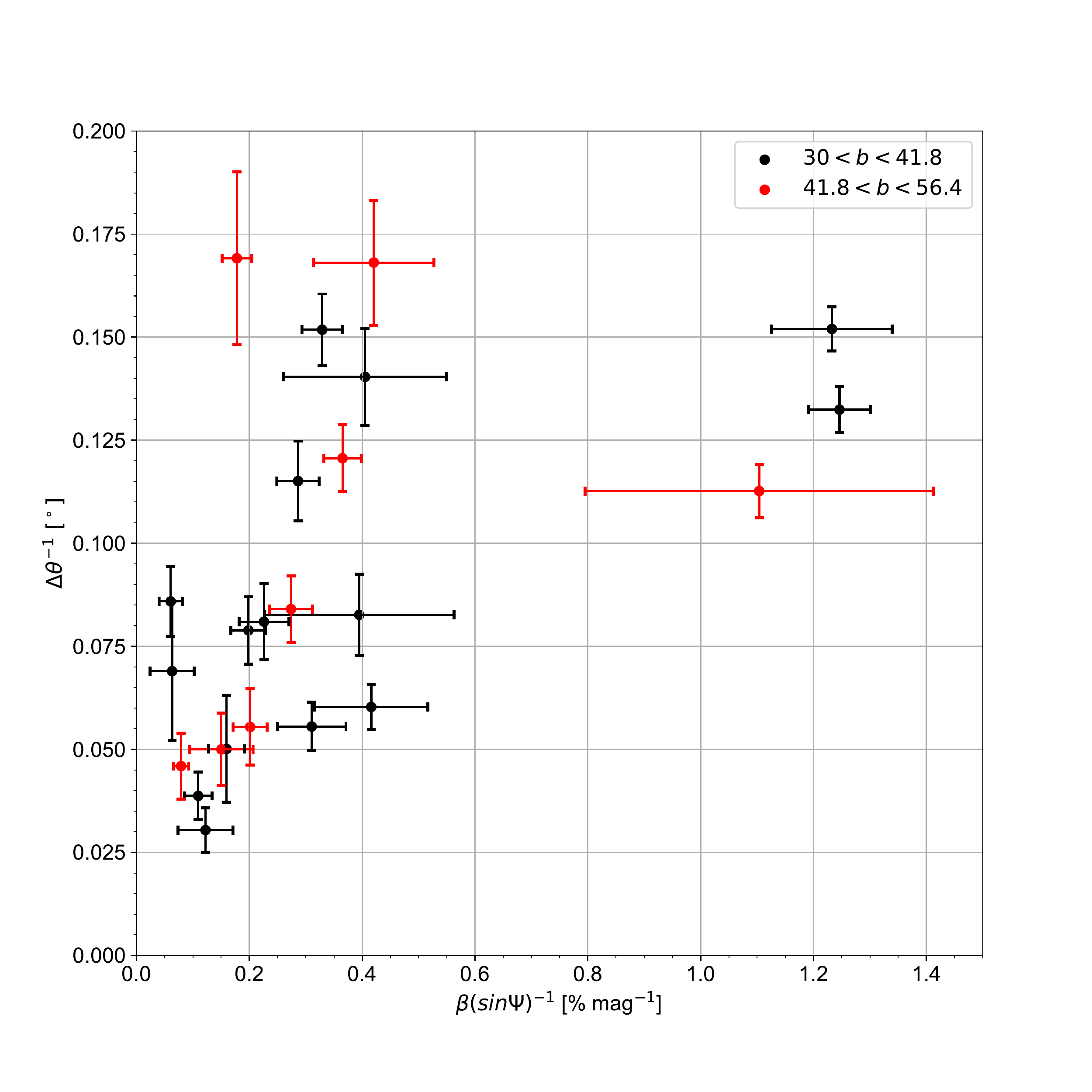}
	\caption{Plot of the inverse of the measured polarization angle dispersion versus the measured $\beta (sin\Psi)^{-1}$. Black data points correspond to $30<b<41.8$ while red data points correspond to $41.8<b<56.4$.}
	\label{pol_disp_beta_compare}
\end{figure}

Comparing the inverse of the angle dispersions with $\beta (sin\Psi)^{-1}$ (Figure \ref{pol_disp_beta_compare}) we find a general correlation for both the low and middle bands.  Spearman rank-order tests support positive correlations with probabilities P$>$99.75\% and P$>$99.87\% for the lower and combined bands, respectively.   For the middle band alone the probability of a positive correlation is limited to  P$>$85\%, but based on only 8 data points.

\section{Discussion} \label{sec:discussion}

\subsection{Grain Alignment} \label{sec:grain_alignment}

Our model of the spatial variations of the fractional polarization in the Local Bubble wall, based on radiation effects, produces over-all very good fits to the data (Table \ref{modelsumtab}, Figures \ref{blow_both_mods}-\ref{pointmodmid}).  This is especially true when focusing on the bluest stars (including the local OB-associations) as the source of the radiation field.  Comparing the fits based on only the light from the OB-associations with those also including field stars, a worsening fit quality is found with later spectral classes added in.  Since the B-parameter in equ. \ref{modeq} is not wavelength dependent in our simple model, the model cannot directly differentiate for the influence of wavelength on the grain alignment.  However, adding a poorly correlated (but spatially isotropic) source to the model will make the total flux closer to a global average in Galactic longitude and ``wash out" any peaks in the correlated terms, as seen in our results.

Radiative Alignment Torque (RAT) theory \citep{lazarian2007} predicts that a dust grain will be aligned only if the impinging radiation has a wavelength less than the grain diameter.  As discussed by e.g. \citet{mathis1977} the total grain-size distribution (aligned and un-aligned) follows a steeply falling size distribution with $n(a)\propto a^{-3.5}$, with a span in sizes of $a<0.01$ to $\sim0.3\mu$m in diffuse gas (where the upper size limit is not well constrained).  \citet{kim1995} showed that for the aligned grains, in diffuse gas, the small size cut-off is $\sim 0.045\mu$m, but otherwise follows the total size distribution.  As discussed e.g. by \citet{bga2015b} this cut-off is consistent with RAT theory and the absence in the general ISM of light short-ward of the Lyman limit.  As shown by \citet{whittet2001} and \citet{bga2007}, as the illuminating light gets progressively redder into clouds the small size cut-off in the aligned grain distribution  grows with extinction.  In an equivalent way, intrinsically redder sources will only be able to align, relatively larger grains.

While the results from \citet{whittet2001} and \citet{bga2007} relied on locating the peak of the polarization curve, through multi-band polarimetry, we can make the plausible argument that the above results are also due to the color dependence of the grain alignment.    Stars are, of course, not monochromatic (nor do they have an infinitely sharp blue cut-off in the Wien tail of their spectral energy distribution), but if we use the peak of the Planck curve, for the effective temperature of a given spectral classes as the short wavelength limit of its light, together with the RAT criterion for alignment of $a>\lambda/2$, we find that for an O5 star (T$_{eff}$=38000K) only grains larger than $\sim 0.04\mu m$ will be aligned.  For an A5 star (T$_{eff}$=8620K) the equivalent size limit is $a>0.17\mu m$.   Given the steep decline in the total grain size distribution, this means that less than 3\% of the grains alignable by an O5 star can be aligned by an A5 star.  In addition, given that the RAT condition is a limit: $a>\lambda/2$, the addition of red light to an existing radiation field containing blue light should be negligible in most circumstances.  Both the dependence of the fractional polarization on the intensity of the (blue) light in the local diffuse radiation field, as well as its implied color dependence, therefore support an origin in radiative grain alignment, and specifically RAT alignment.
 
\subsubsection{The $\alpha$ parameter}\label{alpha}

As briefly discussed in Section \ref{fracpol}, the $\alpha$ value is related to line of sight effects on the fractional polarization, including turbulence.  For a constant grain alignment efficiency along the line of sight, and a fully turbulent medium, $\alpha$ is expected to be $-0.5$.  We find $\langle\alpha\rangle=-0.82\pm0.06$.  A steeper slope than $\alpha=-0.5$ indicates a loss of polarization efficiency with extinction, in addition to the effects of turbulence.  As shown by several authors \citep{alves2014,bga2015b,jones2015}, when the radiation in a cloud has been reddened beyond wavelengths where even the largest grains can be radiatively aligned (cf \citealt{bga2015b}), the slope of the fractional polarization increases to $\alpha$=-1.  The limited column densities of the Local Bubble wall, assuming a ``normal" value of $\langle R_V\rangle=3.1$ - and thus a typical interstellar grain size distribution - would require a somewhat modified explanation here.  This could include a very blue aligning radiation field, which is more rapidly extincted than the diffuse interstellar field.  Another explanation could be that in a medium where the anti-correlation of space density and gas temperature, usually found in the ISM, is not present, the disaligning collisions increase more steeply with extinction than in the general ISM.  Multi-band polarimetry and further density and temperature sensitive data (cf \citet{lehner2003}) are required to address the origin of the large (absolute) value of $\alpha$ and to test these possible alternatives.

\subsection{Magnetic Field Strength} \label{sec:mag_field}

The Davis-Chandrasekhar-Fermi method is based on the fact that, when the equations of motion for a plasma are written down in a magneto-hydrodynamic (MHD) formulation \citep[e.g.][]{choudhuri1998}, the magnetic field lines behave like tensed strings, with the string tension proportional to the field strength.  For a given ``plucking force" of the string, and string mass density, the vibrations of the magnetic field lines can then be derived. This results in the expression \citep{chandrasekhar1953}:
\begin{equation}
\langle B_\perp \rangle^{2} = \frac{4 \pi \rho \Delta v^2_{los}}{\Delta\theta^2}
\label{dcfequ}
\end{equation}  
where $\rho$ is the space density of the material, $\Delta v_{los}$ is the turbulent velocity and $\Delta\theta$ is the position angle dispersion.  If we assume that the magnetic field is frozen into the plasma (i.e. the field drags the plasma along when moving) and that the turbulence is the driving force, we can use the dispersion in the polarization angles to estimate the field strength.

While we do not have any direct probes of the variations of the space density of the gas over our sample, the turbulent velocity of the medium can be estimated from the line widths of resolved absorption lines from the gas.  Under the assumption of a supersonic medium the turbulence in the Local Bubble wall is close to uniform (Figure \ref{turb_plot}) with a possible enhancement in the fourth Galactic quadrant by about a factor 3. However, even disregarding the higher points from the \citet{welty1996} survey in the fourth quadrant, a significant variation exists in the measured position angle dispersion with Galactic longitude, with a notable coherent low value towards the inner Galaxy.

Under the assumption of a constant turbulent velocity, the relative position angle dispersions, normalized to some arbitrary region (here labeled 0), can be written as:
\begin{equation} \label{mag_ratio}
\dfrac{\Delta\theta_i}{\Delta\theta_0} = \sqrt{\dfrac{\rho_i}{\rho_0}}\times \dfrac{\langle B_\perp \rangle_0}{\langle B_\perp \rangle_i}
\end{equation} 
Hence, the systematic variations seen in Figure \ref{glonvspa} could be due to either (or both) variations in the magnetic field strength or the space density.  A smaller angle dispersion (relative to the normative value) requires either a larger magnetic field, or a smaller space density.

\begin{figure}[t]
  	\plotone{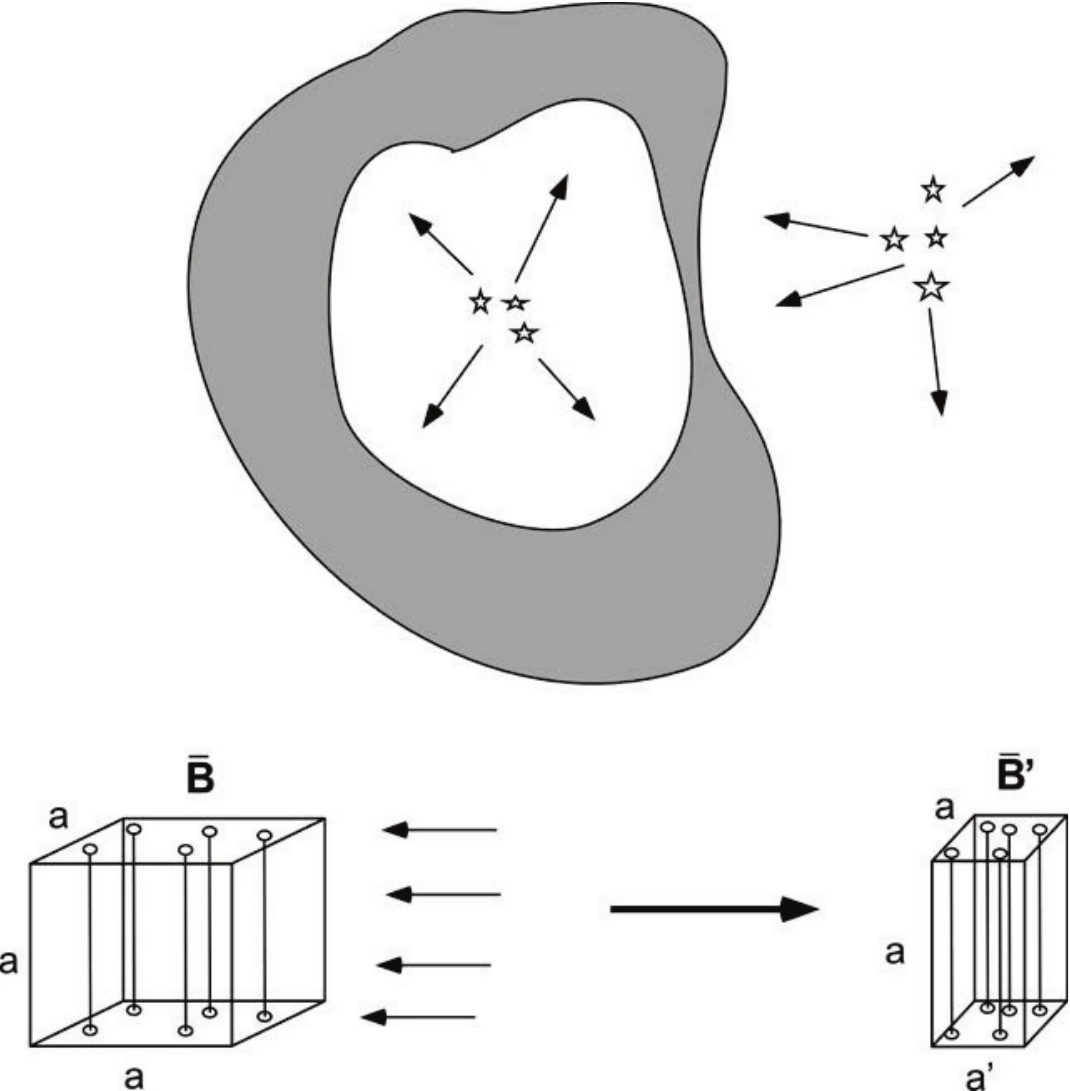}
  	\caption{A sketch of the possible origin of the systematic variations seen in the magnetic pressure, and correlation with the directions of the known, near-by, OB associations.  The Local Bubble is formed by stellar winds and Supernovae remnants from the inside, and the wall (gray) is compressed where this flow is countered by flows from surrounding OB associations (top).  If the magnetic field is frozen in to the plasma, then the component perpendicular to the gas flow directions (thin arrows) is enhanced as the gas is compressed (bottom).}
  	\label{LBWcartoon}
\end{figure}
 
Because we observe a spatial correlation between low position angle dispersion and enhanced radiation field strength, we propose the following hypothesis.  If the radiation field intensity from an OB association is associated with mechanical pressure (e.g. from stellar winds) the bubble wall would be compressed by the countervailing flows from the expansion of the Local Bubble and similar flows from the external OB associations.  If we assume, as indicated by theory \citep{stil2009}, that the magnetic field is mostly parallel to the Local Bubble wall and frozen into the plasma, the field strength would rise as the wall is compressed.  If we consider a simple one-dimensional compression across the average field direction (Figure \ref{LBWcartoon}) the density and field strength enhancements will be linearly correlated.  Because the density enters equation \ref{mag_ratio} as a square root, we would still observe a decreased position \textbf{angle} dispersion.  A more likely scenario might have the gas equilibrate with the surroundings through flows along the field lines (such that $\rho_i = \rho_0$).  This would yield a variation in the angle dispersion directly [inversely] proportional to the magnetic field enhancement. Direct measurements of the gas space density in the two types of regions will be able to probe these scenarios.   As the coverage of the polarization from the Local Bubble expands to include the Galactic plane region and the southern Galactic hemisphere the relationship between local super bubbles and the proposed compression of the Local Bubble wall can also be better tested\textbf{.}
  
\section{Conclusions} \label{sec:conclusions}

We have used large archival databases of optical polarimetry, optical and near-infrared photometry and spectral classification, to study the grain alignment and magnetic field strengths in the wall of the Local Bubble.  We find significant spatial variation both in the grain alignment efficiency, as traced by the fractional polarization (p/A$_V$), and the magnetic field strength, measured through the Davis-Chandrasekhar-Fermi method. Our main results are:
\\
\\
$\bullet$ Assuming the size, shape and mineralogy of the grain distribution, and the gas density are constant throughout the Local Bubble wall, our modeling indicates that the grain alignment, and its variations, is dominated by a radiatively driven mechanism, and that the effects due to the radiation from the nearby OB associations can accurately reproduce the data.  Including a field star sample in the radiation budget, we find that the light from redder stars likely \textbf{does} not contribute significantly to the grain alignment and therefore the fractional polarization.  Given the steep decline in the size distribution of interstellar grains, and the RAT condition for alignment, this result is consistent with RAT theory.  These results support radiatively driven grain alignment and indicate that polarimetry may be able to probe radiation field strength and color variations.  The future inclusion of multi-band polarimetry in this analysis promises to also allow the spectral energy distribution of the radiation field to be examined.
\\
\\
$\bullet$ Assuming a constant space density, the relative magnetic field strength in the Local Bubble wall is found to have systematic variations with one set of values around $8\mu$G (consistent with the earlier estimate of \citet{bga2006}), and others at larger values about three times higher, and up to about 40$\mu$G. We hypothesize that, the correlation between the grain alignment efficiency and magnetic field strength, is due to a compression of the Local Bubble wall caused by the countervailing forces of the outflows generating the Local Bubble and the equivalent ones from the surrounding OB associations, primarily the Sco-Cen associtions.
\\
\\
Even though the \citet{berdyugin2014} survey only covers northern Galactic latitudes of \textit{b}$>30^\circ$, while the centers of the near-by OB associations range from \textit{b}= -20 to +20, the correspondence between the derived alignment-driving sources, the magnetic field strength and the local OB associations is very good.  The ongoing expansion of the polarization survey to lower Galactic latitudes (Berdyugin 2018, private communication) will allow an even more detailed analysis of the grain alignment in the Local Bubble wall.  A quantitative comparison with grain alignment theory is beyond the scope of the present paper, but with the addition of accurate density and temperature data for the Bubble wall gas, these data will allow both further tests of the theory, and several of the uncertain dust grain parameters to be further constrained.  Because of the wavelength dependence of radiative grain alignment, well defined multi-band surveys might additionally allow constraints to be set on the aligning radiation field spectral energy distribution.


\acknowledgments

We gratefully recognize the support by the National Science Foundation through grant 1715867.  The research presented herein has made extensive use of the VizieR catalog access tool at CDS,
Strasbourg, France. The original description of the VizieR service was
published in A\&AS, 143, 23.

We thank Dr. Andrei Berdyugin for generating the polarization catalog and for several helpful discussions.  We thank Dr. Dan Welty for assistance with the high-resolution spectroscopy surveys.

Mr. Medan gratefully acknowledges support from a Fox Fellowship through the Santa Clara University Physics Department, and the support of Prof. Kristin Kulas.



\bibliographystyle{apj}

\cleardoublepage
\begin{deluxetable*}{lllcccccc}
	\tablecaption{Summary Table \label{summarytab}}
	\tablewidth{0pt}
	\tablehead{
		\colhead{Area} & \colhead{\textit{l}} & \colhead{\textit{b}} & \colhead{LBW Angle ($\Psi$)\tablenotemark{$\S$}} & \colhead{$\langle\theta\rangle$\tablenotemark{$\dag$}} & \colhead{$\Delta\theta$\tablenotemark{$\dag$}} & \colhead{$\langle A_v\rangle$\tablenotemark{$\ddag$}} & \colhead{$\alpha$} & \colhead{$\beta$} \\
		\colhead{} & \colhead{[$^\circ$]} & \colhead{[$^\circ$]} & \colhead{[$^\circ$]} & \colhead{[$^\circ$]} & \colhead{[$^\circ$]} & \colhead{[mag]} & \colhead{} & \colhead{[$\%mag^{-1}$]}
	}
	\startdata
	1  & 0-15    & 30-41.8   & $85\pm4$ & $89.4\pm0.3$  & $7.55\pm0.32$  & $0.690\pm0.027$  & $-0.534\pm0.100$  & $1.24\pm0.05$    \\
	2  & 0-15    & 41.8-56.4 & $45\pm2$ & $83.6\pm0.5$ & $5.95\pm0.54$ & $0.532\pm0.024$ & $-0.977\pm0.414$   & $0.295\pm0.074$    \\
	3  & 0-15    & 56.4-90   & \nodata     & $67.4\pm0.6$ & $8.43\pm0.56$ & \nodata          & \nodata           & \nodata            \\
	4  & 15-30   & 30-41.8   & $85\pm4$   & $87.1\pm0.4$ & $6.59\pm0.38$ & $0.446\pm0.023$ & $-0.564\pm0.227$  & $0.328\pm0.036$   \\
	5  & 15-30   & 41.8-56.4 & $82\pm3$   & $91.8\pm0.7$ & $5.91\pm0.73$ & $0.554\pm0.025$ & $-1.11\pm0.23$   & $0.176\pm0.026$   \\
	6  & 15-30   & 56.4-90   & \nodata     & $63.9\pm0.8$ & $12.0\pm0.84$   & \nodata          & \nodata           & \nodata            \\
	7  & 30-45   & 30-41.8   & $80\pm3$   & $97.2\pm1.4$  & $12.3\pm1.4$  & $0.395\pm0.015$ & $-1.26\pm0.18$   & $0.223\pm0.043$   \\
	8  & 30-45   & 41.8-56.4 & $83\pm2$ & $101\pm1$  & $8.29\pm0.56$ & $0.398\pm0.017$ & $-0.547\pm0.140$ & $0.362\pm0.033$   \\
	9  & 30-45   & 56.4-90   & \nodata     & $74.9\pm0.7$ & $8.57\pm0.72$ & \nodata          & \nodata           & \nodata            \\
	10 & 45-60   & 30-41.8   & $75\pm2$ & $110\pm1$   & $11.8\pm1.4$  & $0.273\pm0.029$          & \nodata           & \nodata            \\
	11 & 45-60   & 41.8-56.4 & $87\pm3$   & $100\pm1$  & $7.24\pm0.61$ & $0.487\pm0.030$          & \nodata           & \nodata            \\
	12 & 45-60   & 56.4-90   & \nodata     & $78.5\pm2.9$  & $18.9\pm2.9$  & $0.425\pm0.021$ & $-0.840\pm0.412$   & $0.155\pm0.049$   \\
	13 & 60-75   & 30-41.8   & $84\pm2$ & $110\pm5$   & $20.0\pm5.2$    & $0.360\pm0.024$  & $-0.777\pm0.234$    & $0.159\pm0.032$   \\
	14 & 60-75   & 41.8-56.4 & $86\pm3$ & $103\pm4$   & $21.8\pm3.8$  & $0.523\pm0.027$ & $-1.19\pm0.27$   & $0.0788\pm0.0134$  \\
	15 & 60-75   & 56.4-90   & \nodata     & $82.2\pm8.4$  & $23.2\pm8.4$  & $0.490\pm0.030$          & \nodata           & \nodata            \\
	16 & 75-90   & 30-41.8   & $87\pm2$   & $133\pm5$   & $32.9\pm5.8$  & $0.370\pm0.027$ & $-0.962\pm0.425$   & $0.122\pm0.049$   \\
	17 & 75-90   & 41.8-56.4 & $89\pm4$ & $66.9\pm4.3$  & $21.0\pm4.3$    & $0.495\pm0.036$          & \nodata           & \nodata            \\
	18 & 75-90   & 56.4-90   & \nodata     & \nodata        & \nodata        & \nodata          & \nodata           & \nodata            \\
	19 & 90-105  & 30-41.8   & $83\pm1$   & $154\pm6$   & $26.5\pm6.3$  & $0.334\pm0.026$          & \nodata           & \nodata            \\
	20 & 90-105  & 41.8-56.4 & $89\pm4$ & \nodata & \nodata & $0.375\pm0.032$          & \nodata           & \nodata            \\
	21 & 90-105  & 56.4-90   & $35\pm8$ & $77.9\pm8.8$   & $27.0\pm8.8$    & \nodata          & \nodata           & \nodata            \\
	22 & 105-120 & 30-41.8   & $75\pm3$   & \nodata        & \nodata        & $0.595\pm0.025$ & $-1.02\pm0.18$   & $0.116\pm0.069$   \\
	23 & 105-120 & 41.8-56.4 & $89\pm1$ & \nodata  & \nodata  & $0.388\pm0.029$          & \nodata           & \nodata            \\
	24 & 105-120 & 56.4-90   & $72\pm11$   & \nodata        & \nodata        & \nodata          & \nodata           & \nodata            \\
	25 & 120-135 & 30-41.8   & $69\pm4$ & $20.8\pm3.6$  & $14.5\pm3.6$  & $0.559\pm0.024$ & $-1.33\pm0.23$   & $0.0587\pm0.0362$ \\
	26 & 120-135 & 41.8-56.4 & $79\pm3$   & $54.7\pm5.6$  & $22.5\pm5.6$  & $0.480\pm0.036$          & \nodata           & \nodata            \\
	27 & 120-135 & 56.4-90   & $38\pm8$ & \nodata        & \nodata        & \nodata          & \nodata           & \nodata            \\
	28 & 135-150 & 30-41.8   & $64\pm4$ & $6.64\pm3.84$  & $25.8\pm3.8$  & $0.496\pm0.016$ & $-1.13\pm0.30$    & $0.0979\pm0.0216$    \\
	29 & 135-150 & 41.8-56.4 & $58\pm4$   & $13.5\pm7.5$  & $29.0\pm7.5$    & $0.494\pm0.040$          & \nodata           & \nodata            \\
	30 & 135-150 & 56.4-90   & \nodata     & \nodata        & \nodata        & $0.254\pm0.046$          & \nodata           & \nodata            \\
	31 & 150-165 & 30-41.8   & $54\pm3$   & $4.46\pm1.31$  & $12.7\pm1.3$  & $0.477\pm0.017$ & $-0.828\pm0.214$   & $0.161\pm0.024$   \\
	32 & 150-165 & 41.8-56.4 & $24\pm3$   & $3.17\pm1.19$  & $7.31\pm1.31$  & $0.419\pm0.034$          & \nodata           & \nodata            \\
	33 & 150-165 & 56.4-90   & \nodata     & \nodata        & \nodata        & \nodata          & \nodata           & \nodata            \\
	34 & 165-180 & 30-41.8   & $45\pm2$   & $2.85\pm1.91$  & $18.0\pm1.91$    & $0.361\pm0.020$ & $-0.427\pm0.212$  & $0.220\pm0.042$   \\
	35 & 165-180 & 41.8-56.4 & \nodata     & $166\pm3$   & $17.0\pm3.5$    & $0.284\pm0.029$          & \nodata           & \nodata            \\
	36 & 165-180 & 56.4-90   & \nodata     & \nodata        & \nodata        & \nodata          & \nodata           & \nodata            \\
	37 & 180-195 & 30-41.8   & $55\pm3$ & \nodata        & \nodata        & $0.539\pm0.026$  & $-0.948\pm0.198$   & $0.167\pm0.022$   \\
	38 & 180-195 & 41.8-56.4 & $35\pm3$   & $166\pm3$   & $13.2\pm2.6$  & $0.470\pm0.031$          & \nodata           & \nodata            \\
	39 & 180-195 & 56.4-90   & \nodata     & \nodata        & \nodata        & \nodata          & \nodata           & \nodata            \\
	40 & 195-210 & 30-41.8   & $54\pm5$   & $61.7\pm3.2$  & $18.2\pm3.2$  & $0.456\pm0.038$          & \nodata           & \nodata            \\
	41 & 195-210 & 41.8-56.4 & $35\pm3$   & \nodata        & \nodata        & $0.701\pm0.076$          & \nodata           & \nodata            \\
	42 & 195-210 & 56.4-90   & \nodata     & \nodata        & \nodata        & \nodata          & \nodata           & \nodata            \\
	43 & 210-225 & 30-41.8   & $62\pm4$ & $60.0\pm1.1$    & $11.6\pm1.1$  & $0.420\pm0.026$ & $-1.07\pm0.18$   & $0.0533\pm0.0180$ \\
	44 & 210-225 & 41.8-56.4 & $11\pm2$ & $109\pm4$   & $20.0\pm3.5$    & $0.571\pm0.027$ & $-0.363\pm0.442$  & $0.150\pm0.056$   \\
	45 & 210-225 & 56.4-90   & \nodata     & \nodata        & \nodata        & \nodata          & \nodata           & \nodata            \\
	46 & 225-240 & 30-41.8   & $42\pm2$   & \nodata        & \nodata        & \nodata          & \nodata           & \nodata            \\
	47 & 225-240 & 41.8-56.4 & $11\pm2$ & \nodata        & \nodata        & $0.568\pm0.042$          & \nodata           & \nodata            \\
	48 & 225-240 & 56.4-90   & \nodata     & $46.2\pm1.2$  & $6.46\pm0.92$ & \nodata          & \nodata           & \nodata            \\
	49 & 240-255 & 30-41.8   &$43\pm1$   & \nodata        & \nodata        & $0.392\pm0.042$          & \nodata           & \nodata            \\
	50 & 240-255 & 41.8-56.4 & $44\pm2$ & $64.1\pm6.3$  & $23.6\pm6.3$  & $0.487\pm0.049$          & \nodata           & \nodata            \\
	51 & 240-255 & 56.4-90   & \nodata     & $34.6\pm3.3$  & $14.1\pm3.3$  & \nodata          & \nodata           & \nodata            \\
	52 & 255-270 & 30-41.8   & $83\pm4$ & $84.4\pm3.6$  & $28.2\pm3.6$  & $0.162\pm0.035$          & \nodata           & \nodata            \\
	53 & 255-270 & 41.8-56.4 & $71\pm3$   & $45.8\pm1.0$     & $4.96\pm0.55$ & \nodata          & \nodata           & \nodata            \\
	54 & 255-270 & 56.4-90   & \nodata     & $62.6\pm2.9$  & $19.3\pm2.9$  & \nodata          & \nodata           & \nodata            \\
	55 & 270-285 & 30-41.8   & $55\pm4$ & $43.8\pm1.4$  & $12.1\pm1.4$  & $0.375\pm0.030$ & $-0.527\pm0.440$   & $0.380\pm0.135$   \\
	56 & 270-285 & 41.8-56.4 & $63\pm3$ & $44.5\pm1.2$  & $12.5\pm1.2$  & \nodata          & \nodata           & \nodata            \\
	57 & 270-285 & 56.4-90   & \nodata     & $62.1\pm3.7$  & $14.4\pm3.7$  & \nodata          & \nodata           & \nodata            \\
	58 & 285-300 & 30-41.8   & $54\pm3$ & $70.8\pm3.7$  & $25.2\pm3.7$   & $0.650\pm0.033$          & \nodata           & \nodata            \\
	59 & 285-300 & 41.8-56.4 & $76\pm4$   & $56.0\pm3.0$    & $18.0\pm3.0$    & $0.466\pm0.027$ & $-0.767\pm0.230$   & $0.196\pm0.029$   \\
	60 & 285-300 & 56.4-90   & \nodata     & $68.6\pm0.9$ & $6.81\pm0.86$ & \nodata          & \nodata           & \nodata            \\
	61 & 300-315 & 30-41.8   & $80\pm3$   & $62.1\pm1.5$  & $16.6\pm1.5$  & $0.327\pm0.028$  & $-0.224\pm0.285$  & $0.410\pm0.099$   \\
	62 & 300-315 & 41.8-56.4 & $83\pm3$   & $57.6\pm3.3$  & $16.8\pm3.3$  & $0.281\pm0.032$          & \nodata           & \nodata            \\
	63 & 300-315 & 56.4-90   & \nodata     & $64.0\pm0.8$   & $12.0\pm0.8$   & \nodata          & \nodata           & \nodata            \\
	64 & 315-330 & 30-41.8   & $81\pm3$   & $59.6\pm0.7$ & $8.69\pm0.73$ & $0.419\pm0.023$  & $-0.676\pm0.163$  & $0.283\pm0.037$   \\
	65 & 315-330 & 41.8-56.4 & $79\pm5$   & $62.6\pm1.1$  & $11.9\pm1.1$  & $0.418\pm0.028$ & $-1.21\pm0.18$   & $0.269\pm0.037$   \\
	66 & 315-330 & 56.4-90   & $44\pm5$   & $68.0\pm0.8$   & $10.2\pm0.8$ & \nodata          & \nodata           & \nodata            \\
	67 & 330-345 & 30-41.8   & $73\pm3$   & $66.4\pm0.6$   & $7.12\pm0.60$   & $0.421\pm0.028$ & $-0.820\pm0.453$     & $0.387\pm0.138$  \\
	68 & 330-345 & 41.8-56.4 & $76\pm5$ & $66.3\pm0.7$ & $11.0\pm0.7$   & \nodata          & \nodata           & \nodata            \\
	69 & 330-345 & 56.4-90   & $87\pm5$   & $81.2\pm0.6$ & $9.35\pm0.60$ & \nodata          & \nodata           & \nodata            \\
	70 & 345-360 & 30-41.8   & $78\pm4$   & $79.5\pm0.2$ & $6.58\pm0.23$ & $0.630\pm0.024$  & $-0.599\pm0.191$   & $1.21\pm0.103$      \\
	71 & 345-360 & 41.8-56.4 & $39\pm2$ & $79.8\pm0.5$ & $8.88\pm0.51$ & $0.394\pm0.034$ & $-0.595\pm0.344$    & $0.687\pm0.191$    \\
	72 & 345-360 & 56.4-90   & $43\pm2$   & \nodata        & \nodata        & \nodata          & \nodata           & \nodata           
	\enddata
	\tablenotetext{$\S$}{Areas with no data are where the line of sight along the midpoint does not intersect with the Local Bubble wall. Error estimates are \\based on the change in the wall angle from the midpoint to the edges of the bin. All wall angles are recorded between 0 and 90$^\circ$.}
	\tablenotetext{$\dag$}{Areas with fewer than 10 targets in the \citet{berdyugin2014} catalog, or where the polarization position angles did not yield nor-\\mal distributions, are omitted}
	\tablenotetext{$\ddag$}{Areas for which fewer than 10 targets yielded extinction (or fractional polarization for subsequent columns) estimates of $>2\sigma$ signifi-\\cance are omitted in the A$_V$ and subsequent columns.}
\end{deluxetable*}

\end{document}